\documentclass[twocolumn]{autart}    
\usepackage{graphicx}  
\usepackage{amsmath}
\usepackage{amsfonts} 
\usepackage{appendix}
\usepackage{hyperref}
\usepackage{subfigure}
\usepackage{flushend}

\usepackage{tikz}

\definecolor{region4}{RGB}{238 238 224}
\definecolor{region1}{RGB}{193 255 193}
\definecolor{region3}{RGB}{255 255 224}
\definecolor{region0}{RGB}{245 245 245}
\definecolor{region2}{RGB}{240 248 255}
\definecolor{region5}{RGB}{255 240 245}

\usepackage[ruled,vlined,linesnumbered]{algorithm2e}

\newtheorem{theorem}{Theorem}
\newtheorem{lemma}{Lemma}

\newtheorem{definition}{Definition}
\newtheorem{proposition}{Proposition}
\newtheorem{problem}{Problem}
\newtheorem{remark}{Remark}

\def \AP{\mathcal{AP}}
\def \A{\mathcal{A}}
\def \I{\mathcal{I}}
\def \L{\mathcal{L}}
\def \K{\mathcal{K}}

\def \v{\bar{v}}
\def \x{\bar{x}}

\def \<{\langle}
\def \>{\rangle}
\def \P{\mathcal{P}}
\def \NN{\mathbb{N}}
\def \RR{\mathbb{R}}

\def \pref{\emph{\text{pref}}}
\def \stop{\emph{\text{stop}}}
\def \stra{\textsf{Stra}}
\def \last{\textsf{last}}
\def \reg{\text{reg}}
\def \kw{\textsf{Kw}}
\def \hist{\textsf{Hist}}

\def \succ{\text{Succ}}
\def \trace{\textsf{Trace}}
\def \path{\textsf{Path}}
\def \cost{\textsf{cost}}
\def \play{\textsf{Play}}
\def \S{\mathfrak{S}}
\def \T{\mathbb{T}}

\DeclareMathOperator*{\argmin}{arg\,min}

\begin{document}

\begin{frontmatter}

\title{To Explore or Not to Explore: Regret-based LTL Path Planning in Partially-Known Environments\thanksref{footnoteinfo}} 

\thanks[footnoteinfo]{This paper was presented at IFAC WC, Yokohama, Japan, 2023. Corresponding author: Xiang Yin.}

\author[SJTU]{Jianing Zhao}\ead{jnzhao@sjtu.edu.cn},     
\author[MSU]{Keyi Zhu}\ead{zhukeyi1@msu.edu},
\author[SJTU]{Mingyang Feng}\ead{Fmy-135214@sjtu.edu.cn},
\author[SJTU]{Xiang Yin}\ead{yinxiang@sjtu.edu.cn}  

\address[SJTU]{Department of Automation, Shanghai Jiao Tong University, Shanghai 200240, China}                                             
\address[MSU]{Department of Mechanical Engineering, Michigan State University, East Lansing, MI 48824 USA}


\begin{abstract}                       
In this paper, we investigate the optimal robot path planning problem for high-level specifications described by co-safe linear temporal logic (LTL) formulae.  We consider the scenario where the map geometry of the workspace is \emph{partially-known}. Specifically, we assume that there are some unknown regions, for which the robot does not know their successor regions \emph{a priori} unless it reaches these regions physically. In contrast to the standard game-based approach that optimizes the worst-case cost, in the paper, we propose to use \emph{regret} as a new metric for  planning  in such a partially-known environment. The regret of a plan under a fixed but unknown environment is the difference between the actual cost incurred and the best-response cost the robot could have achieved if it realizes the actual environment with hindsight. We provide an effective algorithm for finding an optimal plan that satisfies the LTL specification while minimizing its regret. A case study on  firefighting  robots is provided to illustrate the proposed framework. We argue that the new metric is more suitable for the scenario of partially-known environment since it captures the trade-off between the actual cost spent and the potential benefit one may obtain for exploring an unknown region. 
\end{abstract}

\end{frontmatter}

\section{Introduction}

Path planning is one of the central problems in autonomous robots. In this context, one needs to design a finite or infinite path for the robot, according to its dynamic and the underlying environment, such that some desired requirements can be fulfilled. In many robotics applications such as search and rescue, persistent surveillance or warehouse delivery, the planning tasks are usually complicated evolving spatial and/or temporal constraints. Therefore,  in the past years,   robot path planning for \emph{high-level specifications} using formal logics has been drawing increasingly more attentions in the literature; see, e.g., \cite{kress2018synthesis,mahulea2020path,kloetzer2020path,yu2022security}. 

Linear temporal logic (LTL) is one of the most popular languages for describing high-level specifications, which supports temporal operators such as ``always", ``eventually" or ``next". In the context of robotic applications, path planning and decision-making for LTL specifications have been investigated very extensively recently. 
For example, \cite{smith2011optimal} studied how to generate an optimal open-loop plan, in the so-called ``prefix-suffix" structure, such that a given LTL formula is fulfilled.
When the results of control actions are non-deterministic, algorithms for synthesizing reactive strategies have been developed using two-player games \cite{fu2016synthesis}. The LTL path planning problem has also been studied for stochastic systems \cite{guo2018probabilistic} to provide probabilistic guarantees and  for multi-robot systems \cite{yu2022distributed} under both global and local tasks.

The aforementioned works on LTL path planning all assume that the environment is known in the sense that the map geometry and the semantic structure are both available at the planning stage. In practice, however, the environment may be partially-known such that the robot needs to explore the map geometry as well as the region semantics on-the-fly. To this end, in \cite{guo2015multi}, the authors provided a re-planning algorithm based on the system model updated online.  In  \cite{lahijanian2016iterative}, the authors proposed an iterative planning algorithm in uncertain environments where unknown obstacles may appear. A learning-based algorithm is proposed in \cite{cai2021learning} for LTL planning in stochastic environments with unknown transition probabilities. Recently, \cite{kantaros2022perception} investigated the LTL planning problem under environments with known map geometries  but with semantic uncertainties.

In this paper, we also investigate the LTL path planning for robots in \emph{partially-known} environments. Specifically, here we assume that the location of each region in the map is perfectly known but, for some regions, the robot does not know their successor regions \emph{a priori} unless it reaches these regions physically. For example, in Figure~1, the dashed line between regions $2$ and $5$ denotes a possible wall that may prevent the robot from reaching region $5$ directly from region $2$. Initially, the robot knows the possibility of the wall, but it will actually know the (non-)existence of the wall only when reaching region $2$.  Here, we distinguish between the terminologies of \emph{non-determinsitic} environments and \emph{partially-known} environments. Specifically, the former is referred to the scenario where the outcome of the environment is purely random in the sense that even for the same visit, the environment may behave differently. However, the partially-known environment is referred to the case where the robot has \emph{information uncertainty} regarding the true world initially, but the underlying actual environment is still fixed and deterministic.

To solve the path planning problem in partially-known environments, a direct approach is to follow the same idea for planning in non-deterministic environments, where game-based approaches are usually used to minimize the \emph{worst-case cost}. Still, let us consider Figure~1, where the robot aims to reach target region $5$ with shortest distance. 
Using a worst-case-based approach, the robot will follow the red trajectory. This is because the short-cut from regions $2$ to $5$ may not exist; if it goes to region $2$, then in the worst-case, it will spend additional effort to go back. However, by taking the red trajectory, the robot may \emph{heavily regret} by thinking that it should have taken the short-cut at region $2$ if it knows \emph{with hindsight} that the wall does not exist. Therefore, a more natural and human-like plan  is to first go to region $2$ to take a look at whether there is a wall. If not, then it can take the short-cut, which saves 7 units cost.  Otherwise, the robot needs to go back to the red trajectory.  Compared with the red path, although this approach may have two more units cost than the worst-case, it takes the potential huge advantage of exploring the unknown regions. 

\newcommand\kk{1.2}
\begin{figure}
\centering
\begin{tikzpicture}  
	\draw[fill=region0,draw=white] (0/\kk,4/\kk)--(0/\kk,2/\kk)--(1/\kk,2/\kk)--(1/\kk,4/\kk);
	\draw[fill=region5,draw=white] (1/\kk,2/\kk)--(3/\kk,2/\kk)--(3/\kk,3/\kk)--(1/\kk,3/\kk);
	\draw[fill=region1,draw=white] (0/\kk,1/\kk)--(1/\kk,1/\kk)--(1/\kk,2/\kk)--(0/\kk,2/\kk);
	\draw[fill=region3,draw=white] (0/\kk,0/\kk)--(4/\kk,0/\kk)--(4/\kk,1/\kk)--(0/\kk,1/\kk);
	\draw[fill=region4,draw=white] (3/\kk,1/\kk)--(4/\kk,1/\kk)--(4/\kk,4/\kk)--(1/\kk,4/\kk)--(1/\kk,3/\kk)--(3/\kk,3/\kk)--(3/\kk,1/\kk);
	\draw[fill=region2,draw=white] (1/\kk,1/\kk)--(3/\kk,1/\kk)--(3/\kk,2/\kk)--(1/\kk,2/\kk)--(1/\kk,1/\kk);
	
	\node[] [xshift=0.5cm/\kk, yshift=3cm/\kk] (0label) {$0$};
	\node[] [xshift=0.5cm/\kk, yshift=1.5cm/\kk] (1label) { $1$};
	\node[] [xshift=2cm/\kk, yshift=1.5cm/\kk] (2label) {$2$};
	\node[] [xshift=2cm/\kk, yshift=0.22cm/\kk] (3label) {$3$};
	\node[] [xshift=2cm/\kk, yshift=3.5cm/\kk] (4label) {$4$};
	\node[] [xshift=2cm/\kk, yshift=2.5cm/\kk] (5label) {$5$};
	
	\draw[fill=blue,draw=blue] (1.5/\kk,1.66/\kk) circle (2pt);
	\draw[fill=blue,draw=blue] (0.66/\kk,2.5/\kk) circle (2pt);
	\draw[very thick, draw=blue] (0.66/\kk,2.5/\kk)--(0.66/\kk,1.66/\kk)--(1.5/\kk,1.66/\kk);
	\draw[-stealth, very thick, draw=blue] (1.5/\kk,1.66/\kk)--(1.5/\kk,1.33/\kk)--(0.66/\kk,1.33/\kk)--(0.66/\kk,0.5/\kk);
	\draw[-stealth, very thick, dashed, draw=blue] (1.5/\kk,1.66/\kk)--(1.5/\kk,2.5/\kk);
	\draw[-stealth, very thick, draw=blue] (0.66/\kk,0.55/\kk)--(3.45/\kk,0.55/\kk)--(3.45/\kk,3.45/\kk)--(2.55/\kk,3.45/\kk)--(2.55/\kk,2.5/\kk);
	
	\draw[fill=red,draw=red] (0.33/\kk,2.5/\kk) circle (2pt);
	\draw[-stealth, very thick, draw=red] (0.33/\kk,2.5/\kk)--(0.33/\kk,0.45/\kk)--(3.55/\kk,0.45/\kk)--(3.55/\kk,3.55/\kk)--(2.45/\kk,3.55/\kk)--(2.45/\kk,2.5/\kk);
	
	\foreach \x in {0,1,...,4}
	\draw[densely dotted] (\x/\kk, 0/\kk)--(\x/\kk, 4/\kk);
	\foreach \y in {0,1,...,4}
	\draw[densely dotted] (0/\kk, \y/\kk)--(4/\kk, \y/\kk);
	\draw[ultra thick] (1/\kk, 3/\kk)--(2/\kk, 3/\kk);
	\draw[ultra thick] (1/\kk, 4/\kk)--(1/\kk, 2/\kk);
	\draw[ultra thick] (1/\kk, 1/\kk)--(3/\kk, 1/\kk);
	\draw[ultra thick] (3/\kk, 3/\kk)--(3/\kk, 1/\kk);
	\draw[ultra thick] (2/\kk, 2/\kk)--(3/\kk, 2/\kk);
	\draw[ultra thick, dotted] (1/\kk, 2/\kk)--(2/\kk, 2/\kk);
	\draw[ultra thick] (0/\kk, 0/\kk)--(4/\kk, 0/\kk)--(4/\kk, 4/\kk)--(0/\kk,4/\kk)--(0/\kk,0/\kk);
	
	\draw[ultra thick, dashed] (4.4/\kk, 2.2/\kk)--(6/\kk,2.2/\kk);
	\node[] [xshift=5.3cm/\kk, yshift=2.55cm/\kk] (1label) {\footnotesize Possible Wall};
	\draw[ultra thick] (4.4/\kk, 3/\kk)--(6/\kk,3/\kk);
	\node[] [xshift=5.22cm/\kk, yshift=3.35cm/\kk] (1label) {\footnotesize Known Wall};
	
	\draw[-stealth, very thick, draw=blue, dashed] (5.6/\kk, 1.2/\kk)--(6.8/\kk,1.2/\kk);
	\draw[-stealth, very thick, draw=blue] (4.4/\kk, 1.2/\kk)--(5.6/\kk,1.2/\kk);
	\node[] [xshift=5.68cm/\kk, yshift=1.55cm/\kk] (1label) {\footnotesize Regret-Based Plan};
	\draw[-stealth, very thick, draw=red] (4.4/\kk, 0.4/\kk)--(6.8/\kk,0.4/\kk);
	\node[] [xshift=6.01cm/\kk, yshift=0.75cm/\kk] (5label) {\footnotesize Worst-Case-Based Plan};
	\end{tikzpicture}
	\caption{A motivating example, where a robot needs to reach region $5$ from regin $0$ with partially-known environment information.}
	\label{fig:motiexam}
\end{figure}
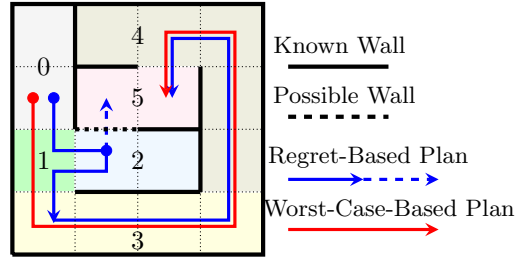 

In this paper, we formulate and solve a new type of LTL optimal path planning problem for robots working in an aforementioned partially-known environment. We adopt the notion of \emph{regret} from game theory \cite{yannakakis2018artificial} as the optimality metric. We propose the structure of \emph{partially-known weighted transition systems} (PK-WTS) as the model that contains the set of all possible actual environments. The regret of a plan under a fixed but unknown environment is defined as the difference between its actual cost and the best-response cost it could have achieved after it knows the actual environment with hindsight. A value iteration algorithm is developed for computing an optimal  strategy such that (i) it satisfies the LTL requirement under any possible environment; and (ii) minimizes its regret. 
We illustrate by case studies that, compared with the worst-case-based synthesis for non-deterministic environments, the proposed regret-based synthesis is more suitable for partially-known environments. 

The regret minimization problem is an emerging topic in the context of graph games; see, e.g.,  \cite{filiot2010iterated,hunter2017reactive,cadilhac2019impatient}. 
Particularly, \cite{filiot2010iterated} is most related to our problem setting, where it solves a reachability game with minimal regret by a graph-unfolding algorithm. 
However, such a graph-unfolding approach is unnecessary when the regret-minimizing strategy is synthesized on a game arena that reflects the the map geometry of the environment to explore rather than a general bipartite graph. 
Instead, we propose a more efficient algorithm, based on a new weight function construction, to solve the regret-minimizing exploration problem.
We reduce the computation complexity from \emph{pseudo-polynomial} in \cite{filiot2010iterated} to \emph{polynomial}.
In the context of robotic applications, the recent work \cite{muvvala2022let} uses regret to optimize  human-robot collaboration strategies, based on the algorithm in \cite{filiot2010iterated}, while the issue of exploration is still not handled.  In this work, we use regret to capture the issue of exploration in partially-known environments, which is also different from the purpose of \cite{muvvala2022let}.

The remaining part of this paper is organized as follows: 
In Section~\ref{sec:fully}, we review some necessary preliminaries and the standard LTL planning in  fully-known environments. In Section~\ref{sec:parti}, we present the mathematical model for partially-known environments and introduce regret as the performance metric. In Section~\ref{sec:KGA}, we transfer the planning problem as a two-player game on a new structure named \emph{knowledge-based game arena}.
In Section~\ref{sec:algorithm}, we propose an efficient algorithm to solve the regret-minimizing strategy synthesis problem.
Numerical simulation and a case study on firefighting robot is provided in Section~\ref{sec:case} to show the effectiveness and the performance of the regret-based strategy compared to the other exploration strategies as well as the algorithm in \cite{filiot2010iterated}.  Finally, we conclude the paper in Section~\ref{sec:conclusion}.

\section{LTL Planning in Fully-Known Environments}\label{sec:fully}
In this section, we briefly review some necessary preliminaries and the standard approach for solving the LTL planning problem in a fully-known environment. 

\subsection{Weighted Transition Systems}
When the environment of the workspace is fully-known, the mobility of the agent (or map geometry) is usually modeled as a \emph{weighted transition system} (WTS)
\[
T=(X,x_0,\delta_T,w_T,\AP,L), 
\]
where $X$ is a set of states representing different regions of the workspace;
$x_0\!\in\! X$ is the initial state representing the starting region of the agent; $\delta_T\!:\!X\!\to\! 2^X$ is the transition function  such that, starting from each state $x\!\in\! X$, the agent can move directly to any of its successor state $x'\!\in\!\delta_T(x)$. We also refer $\delta_T(x)$ to as the \emph{successor states} of $x$; 
$w\!:\!X\times X\!\to\!\RR$ is a cost function such that $w(x,x')$ represents the cost incurred when the agent moves from $x$ to $x'$; 
$\AP$ is the set of atomic propositions; and
$L\!:\!X\!\to\!2^\AP$ is a labeling function assigning each state a set of atomic propositions.

Given a WTS $T$, an infinite \emph{path} of $T$ is an infinite sequence of states 
$\rho\!=\!x_0x_1x_2\cdots \!\in\! X^\omega$ 
such that $x_{i+1}\!\in\!\delta_T(x_i), i\geq 0$. 
A finite path is defined analogously. 
We denote by $\path^\omega(T)$ and $\path^*(T)$ the sets of all infinite paths and finite paths in $T$, respectively.  
Given a finite path $\rho\!=\!x_0x_1\cdots x_n\!\in\!\path^*(T)$, its cost is defined as the sum of all transition weights in it, which is denoted by 
$\cost(\rho)\!=\!\sum_{i=0}^{n-1}w(x_i,x_{i+1})$. 
The \emph{trace} of  an infinite path $\rho\!=\!x_0x_1x_2\cdots \!\in\! X^\omega$ is an infinite sequence over $2^\AP$ denoted by $L(\rho)\!=\!L(x_0)L(x_1)\cdots$. Analogously, we denote by $\trace^\omega(T)$ and $\trace^*(T)$ the sets of all infinite traces and finite traces in $T$, respectively.

\subsection{Linear Temporal Logic Specifications}
The syntax of general LTL formula is given as follows
\[
\phi=\top\mid a \mid\neg \phi \mid\phi_1\wedge\phi_2 \mid\bigcirc\phi\mid\phi_1 U \phi_2, 
\]
where 
$\top$ stands for the ``true" predicate; 
$a\!\in\!\mathcal{AP}$ is an atomic proposition; 
$\neg$ and $\wedge$ are Boolean operators ``negation" and ``conjunction", respectively; 
$\bigcirc$ and $U$ denote temporal operators ``next" and ``until", respectively. 
One can also derive other temporal operators such as 
``eventually"  by $\lozenge \phi \!=\!\top U \phi$.   
LTL formulae are evaluated over infinite words; the readers are referred to \cite{baier2008principles} for the semantics of LTL.   
Specifically, an infinite word $\tau\!\in\! (2^{\mathcal{AP}})^\omega$ is an infinite sequence over alphabet $2^{\mathcal{AP}}$. 
We write $\tau\!\models\! \phi$ if $\tau$ satisfies LTL formula $\phi$.

In this paper, we focus on a widely used fragment of LTL formulae called the \emph{co-safe LTL} (scLTL) formulae.   
Specifically, an scLTL formula requires that  the negation operator $\neg$ can only be applied in front of atomic propositions. 
Consequently, one cannot use ``always" $\square$ in scLTL. 
Although the semantics of LTL are defined over infinite words,  it is well-known that any infinite word satisfying a co-safe LTL formula has a \emph{finite good prefix}.   
Specifically,  a good prefix is a finite word $\tau'=\tau_1\cdots \tau_n\in (2^{\mathcal{AP}})^*$ such that $\tau'\tau''\models\phi$ for any $\tau'' \in (2^{\mathcal{AP}})^\omega$.   
We denote by $\L_{\pref}^\phi$ the set of all finite good prefixes of scLTL formula $\phi$.

For any scLTL formula $\phi$, its good prefixes $\L_{\pref}^\phi$ can be accepted by a \emph{deterministic finite automaton} (DFA). Formally, a DFA is a 5-tuple
$\A=(Q,q_0,\Sigma,f,Q_F)$, where
$Q$ is the set of states; $q_0\in Q$ is the initial state; $\Sigma$ is the alphabet;
$f\!:\!Q\times \Sigma\!\to\! Q$ is a transition function; and $Q_F\subseteq Q$ is the set of accepting states. 
The transition function can also be extended  to $f\!:\!Q\times \Sigma^*\!\to\! Q$ recursively. 
A finite word $\tau\in \Sigma^*$ is said to be \emph{accepted} by $\A$  if $f(q_0,\tau)\in Q_F$; we denote by $\L(\A)$ the set of all accepted words. 
Then for any scLTL formula $\phi$ defined over $\AP$,   we can always build a DFA over alphabet $\Sigma=2^{\AP}$, denoted by  $\A_\phi\!=\!(Q,q_0,2^{\AP},f,Q_F)$, such that $\L(\A_\phi)\!=\!\L_\pref^\phi$.

\subsection{Path Planning for scLTL Specifications}\label{subsec-product}
Given a WTS $T$ and an scLTL formula $\phi$, the path planning problem is to find an finite path (a.k.a.\ a plan) $\rho\in \path^*(T)$ such that 
$L(\rho) \in \L_\pref^\phi$ and, at the same time, its cost $\cost(\rho)$ is minimized.  

To solve the scLTL planning problem, the standard approach is to build the \emph{product system} between  WTS $T\!=\!(X,x_0,\delta_T,w,\AP,L)$ and DFA $\A_\phi\!=\!(Q,q_0,\Sigma,f,Q_F)$, which is a new (unlabeled) WTS
\[
P=T\otimes\A_\phi=(S,s_0,\delta_P,w_P,S_F), 
\]
where 
  $S\!=\!X\!\times\! Q$ is the set of states;
  $s_0\!=\!(x_0,q_0)$ is the initial state;
  $\delta_P\!:\!S \!\to\! 2^S$ is the transition function defined by:  
  for any $s\!=\!(x,q)\!\in\! S$,  we have
  $\delta_P(  s )\!=\!\{    (x',q')\!\in\! S \mid  x'\!\in\!\delta_T(x)\wedge q'\!=\! f(q,L(x)) \}$;
  $w_P\!:\! S\!\times\! S\!\to\!\RR$ is the weight function defined by: 
  for any $s\!=\!(x,q),s'\!=\!(x',q')\!\in\! S$, we have $w_P( s,s' )\!=\!w(x,x')$; and
  $S_F\!=\!X\times Q_F$ is the set of accepting states. 
By construction, 
for any path $\rho\!=\!(x_0,q_0)\cdots(x_n,q_n)$ in the product system,  
$(x_n,q_n)\!\in\! S_F$ implies  $\rho\!=\!x_0\cdots x_n\!\in\! \path^*(T)$  and  $L(\rho)\!\in\! \L_{\pref}^\phi$. 
Therefore, to solve the scLTL planning  problem, it suffices to find a path with minimum weight from the initial state to accepting states $S_F$ in the product system.  

\section{Planning in Partially-Known Environments}\label{sec:parti}
The above reviewed shortest-path-search-based LTL planning method crucially  depends on that the mobility of the robot, or the environment map $T$ is perfectly known. This method, however, is not suitable for the case of  \emph{partially-known} environments. 
To be specific,  we consider a partially-known environment in the following setting:
\begin{itemize}
    \item[A1] 
    The agent knows the existence of all regions in the environment as well as their semantics (atomic propositions hold at each region); 
    \item[A2] 
    The successor regions of each region are fixed, but the agent may not know, \emph{a priori}, what are the actual successor regions it can move to;
    \item[A3]
    Once the agent physically reaches a region, it will know the successor regions of this region precisely.  
\end{itemize}

In this section, we will   provide a formal model for such a partially-known environment using the new structure of \emph{partially-known weighted transition systems} and use \emph{regret} as a new metric for evaluating the performance of the  agent's plan in a partially-known environment.

\subsection{Partially-Known Weighted Transition Systems}

\begin{definition}[Partially-Known WTS]
A partially-known weighted transition system (PK-WTS) is a 6-tuple 
\[
\T=(X,x_0,\Delta,w,\AP,L), 
\]
where, similar  to a WTS, 
$X$ is the set of states with initial state $x_0\!\in\! X$, 
$w\!:\!X\times X\!\to\!\RR$ is the cost function 
and $L\!:\!X\!\to\! 2^\AP$ is a labeling function that assigns each state a set of atomic propositions.  
Different from the WTS, 
\[
\Delta:X\to 2^{2^X}\]
is called a \emph{successor-pattern function} that assigns each state $x\in X$ a family of successor states.  \vspace{3pt}
\end{definition} 

The intuition of the PK-WTS $\T$ is explained as follows. 
Essentially, PK-WTS is used to describe the \emph{possible world} from the perspective of the agent. Specifically, under  assumptions A1-A3, the agent has some prior information regarding the successor states of each unknown region but does not know which one is true before it actually visits the region. Therefore, in PK-WTS $\T$, for each state $x\!\in\! X$,  we have $\Delta(x)=\{o_1,\dots,o_{|\Delta(x)|}\}$, where
each $o_i\in 2^X$ is called a \emph{successor-pattern} representing a possible  set of actual successor states at state $x$.  
Hereafter, we will also refer each $o_i\in \Delta(x)$ to as an \emph{observation} at state $x$ since the agent ``observes" its successor states when exploring state $x$. 
Therefore, for each state $x\in X$, we say $x$ is a
\begin{itemize}
    \item 
    \emph{known state}  if $|\Delta(x)|=1$; and 
    \item
    \emph{unknown state}  if $|\Delta(x)|>1$. 
\end{itemize}
We assume that the initial state $x_0$ is known since the agent has already stayed at $x_0$ so that it has the precise information regarding the successor states of $x_0$. 
Therefore, we can partition the state space as 
$X=X_{kno}\dot{\cup}X_{un}$,
where $X_{kno}$ is the set of known states and $X_{un}$ is the set of unknown states.

In reality, the agent is moving in a specific environment that is compatible with the possible world $\T$, although itself does not know this a priori. 
Formally,  we say a WTS $T=(X,x_0,\delta_T,w,\AP,L)$ is \emph{compatible} with PK-WTS  $\T$, denoted by $T\in\T$, if  $\forall x\in X: \delta_T(x)\in\Delta(x)$. 
Clearly, if all states in $\T$ are known, then its compatible WTS is unique.

\subsection{History and Knowledge Updates}
In the partially-known setting, the agent cannot make decision only based on the finite sequence of states it has visited. In addition, it should also consider what it observed (successor-pattern) at each state visited. Note that, when the agent visits a known state $x\in X_{kno}$, it will not gain any useful information about the environment since $\Delta(x)$ is already a singleton.  Only when  the agent visits an unknown state, it will gain new information and successor-pattern at this state will become known from then on. Therefore, we refer the visit to an unknown state to as an \emph{exploration}. 

To capture the result of an exploration, 
we call a tuple $\kappa\!=\!(x,o)\!\in\! X\times 2^X$, where $o \in\Delta(x)$, 
a \emph{knowledge} obtained when exploring state $x$, which means the agent knows that the successor states of $x$ are $o$.   
For each knowledge $\kappa$, we denote by $\kappa(x)$ and $\kappa(o)$, respectively, its first and second components, i.e., $\kappa=(\kappa(x),\kappa(o))$. 
We denote by 
\begin{equation}\label{eq:know-domain}
\textsf{Kw}=\{  \kappa \in X\times 2^X \mid  \kappa(o) \in\Delta(\kappa(x))\}, 
\end{equation}
the set of all possible knowledges. 

A \emph{history} in $\T$ is a finite sequence of knowledges  
\begin{equation}\label{eq:hist}
\hbar= \kappa_0\kappa_1\cdots\kappa_n=  (x_0,o_0)(x_1,o_1)\cdots(x_n,o_n) \in \textsf{Kw}^*
\end{equation}
such that 
\begin{enumerate} 
    \item 
    for any $i=0,\dots, n-1$, we have $x_{i+1}\in o_i$; and 
    \item 
    for any $i,j=1,\dots,n$, we have $x_i=x_j\Rightarrow o_i=o_j$.
\end{enumerate}
Intuitively, the first condition says that the agent can only go to one of its actual successor states in $o_i$. 
The second condition captures the fact that the actual environment is partially-known but \emph{fixed}; hence, the agent will observe the same successor-pattern for different visits of the same state. 
For history $\hbar\!=\!\kappa_0\kappa_1\cdots\kappa_n\!\in\!\textsf{Kw}^*$, we call $\kappa_0(x)\kappa_1(x)\cdots\kappa_n(x)$ its path. 
We denote by $\path^*(\T)$ and $\textsf{Hist}^*(\T)$ the set of all paths and histories generated in PK-WTS $\T$, respectively.

Along each history $\hbar\!\in\!\textsf{Kw}^*$, we denote by an \emph{ordered set} $\K_\hbar\subseteq\textsf{Kw}^*$ the knowledge-set the agent has obtained.
Given a knowledge-set $\K$, we say a state $x\in X$ has been explored in $\K$, if $(x,o)\in\K$ for some $o$;
we denote by $X(\K)$ the set of explored states in $\K$. 
Then, we define the updates of the knowledge-set $\K$ as follows:
\begin{itemize}
    \item for $\hbar=\kappa_0$, we define
    \begin{equation}\label{eq-initialknowledge}
        \K_0:=\K_\hbar=\<(x,o)\mid x\in X_{kno},\Delta(x)=\{o\}\>
    \end{equation}
    where the order of each $(x,o)$ is arbitrary;
    \item for each $\hbar'\!=\!\hbar\kappa$ with $\K_\hbar\!=\!\<\kappa_1,\ldots,\kappa_{|\K_\hbar|}\>$, we define
    \begin{equation}
        \K_{\hbar'}=\begin{cases} 
        \K_\hbar, &\text{ if }\kappa(x)\in X(\K_\hbar)\\
        \<\kappa_1,\ldots,\kappa_{|\K_\hbar|},\kappa\>, &\text{ otherwise.}
        \end{cases}
    \end{equation}
\end{itemize}

For convenience, we define 
\begin{equation}\label{eq:knowset}
\mathbb{KW}\!=\!\{  \K\!\in\!   2^{\textsf{Kw}} \!\mid\!  \forall \kappa,\kappa'\!\in\! \K\!:\!  \kappa(x)\!=\! \kappa'(x) \!\Rightarrow\! \kappa(o)\!=\! \kappa'(o)           \} 
\end{equation}
as the set of all knowledge-sets.
Given a knowledge-set $\K\!\in\!\mathbb{KW}$,
if $x\!\in\! X(\K)$,
we denote by $o_\K(x)\!\in\! 2^X$ the unique observation which satisfies
$(x,o_\K(x)) \!\in\! \K$. Moreover, given two knowledge-sets $\K,\K'\!\in\!\mathbb{KW}$ with $\K\!=\!\<\kappa_1,\ldots,\kappa_{|\K|}\>$ and a new knowledge $\kappa\!\in\!\textsf{Kw}$, we define the knowledge update function as
\begin{equation}
    \K'=\textsf{update}(\K,\kappa)
\end{equation}
satisfying
\begin{equation}\label{eq-update}
    \K'=\begin{cases}
    \K,&\text{ if }\kappa(x)\in X(\K)\\
    \<\kappa_1,\ldots,\kappa_{|\K|},\kappa\>,&\text{ otherwise.}
    \end{cases}\notag
\end{equation}

%In the above definition of knowledge-set, we use the ordered set to record not only the knowledges the agent has obtained, but also the sequence of exploring the unknown states.
%For two knowledge-sets $\K_1,\K_2$, we say they are component-equivalent, denoted by
%$\K_1\simeq\K_2$,
%if all their components are the same, i.e., i) $\forall\kappa\in\K_1:\kappa\in\K_2$ and 
%ii) $\forall\kappa\in\K_2:\kappa\in\K_1$.

With an updated knowledge-set, the agent can maintain a finer possible world by incorporating with the knowledges it obtained. Specifically, 
by having knowledge-set $\K \in \mathbb{KW}$, the agent can update the PK-WTS 
$\T=(X,x_0,\Delta,w,\AP,L)$  to a finer PK-WTS 
\begin{equation}\label{eq:refine}
    \T'=\textsf{refine}(\T,\K)\!=\!(X,x_0,\Delta',w,\AP,L)
\end{equation}
where  for any $x\in X$, we have
\[
\Delta'(x) = 
\left\{ \begin{array}{ll}
\{ o_\K(x)   \} & \text{if }  x\in X(\K) \\
\Delta(x) & \text{if }  x\notin X(\K) 
\end{array} 
\right..
\]
Note that, the above update function is well-defined since 
conflict knowledges $\<x,o\>,\<x,o'\>\in \textsf{Kw}$ such that  $o\neq o'$ cannot belong to the same knowledge-set by the definition of history and Equation~\eqref{eq:knowset}.

\subsection{Strategy and Regret}\label{subset-regretminmax}
Under the setting of partially-known environment, the plan is no longer an open-loop sequence. Instead, it is a \emph{strategy} that determines the next state the agent should go to based what has been visited and what has known, which are environment dependent. Formally, a strategy is a function
$\xi:\hist^*(\T)\to X\cup \{\stop\}$
such that  
for any $\hbar= \kappa_1 \cdots \kappa_n$,  where $\kappa_i=\<x_i,o_i\>$, 
either 
(i)  $\xi(\hbar)\in o_n$, i.e., it decides to move to some successor state;  or (ii) $\xi(\hbar)=\stop$, i.e., the plan is terminated.  
We denote by $\stra(\T)$ the set of all strategies for $\T$.

Although a strategy is designed to handle all possible actual environments in the possible world $\T$, when it is applied to an actual environment $T\in \T$, the outcome of the strategy can be completely determined. 
We denote by 
$\rho^{T}_\xi\!=\!x_0x_1\cdots x_n\!\in\! X^*$
the finite path induced by strategy $\xi$ in environment $T\in \T$, which is the unique path such that 
\begin{itemize}
    \item 
     $\forall i< n:\xi( \<x_0,\delta_T(x_0)\>\cdots    \<x_i,\delta_T(x_i)\>    )=x_{i+1}$; 
    and 
    \item
    $\xi( \<x_0,\delta_T(x_0)\>\cdots    \<x_n,\delta_T(x_n)\>    )=\stop$.
\end{itemize}
Note that the agent does not know \emph{a priori} which $T\in \T$ is the actual environment. 
To guarantee the accomplishment of the LTL task, a strategy $\xi$ should satisfy  
\begin{equation}\label{eq:LTL-sa}
\forall T\in \T:  \rho^{T}_\xi\in \L_{\pref}^\phi. 
\end{equation}
We denote by $\stra_\phi(\T)$   all strategies satisfying~\eqref{eq:LTL-sa}.

To evaluate the performance of strategy $\xi$, one approach is to consider the \emph{worst-case cost} of the strategy among all possible environment, i.e.,
\begin{equation}\label{eq:worstcase}
    \textsf{cost}_{\text{\emph{worst}}}(\xi):=\max_{T\in \mathbb{T}}\cost(\rho^T_\xi)
\end{equation}
However, as we have illustrated by the example in Figure~1, this metric cannot capture the potential benefit obtained from exploring unknown states and the agent may regret due to the unexploration.  To capture this issue, in this work, we propose the notion of \emph{regret} as the metric to evaluate the performance of a strategy.  

\begin{definition}[Regret]\label{def:regret-in-T}
Given a partially-known environment described by PK-WTS $\T$ and a task described by an scLTL $\phi$,  the \emph{regret} of strategy $\xi$ is defined by
\begin{equation}
    \reg_\T(\xi)=\max_{T\in\T}\left(\cost(\rho_{\xi}^T)-\min_{\xi'\in\stra_\phi(\T)}\cost(\rho_{\xi'}^T)\right)
\end{equation}\vspace{0pt} 
\end{definition}

The intuition of the above notion of regret is explained as follows. 
For each strategy $\xi$ and each actual environment $T$, 
$\cost(\rho_{\xi}^T)$ is the actual cost incurred when applying this strategy to this specific environment, while $\min_{\xi'\in\stra_\phi(\T)}\cost(\rho_{\xi'}^T)$ is  cost of the best-response strategy the agent should have  taken if it knows the actual environment $T$  with hindsight. 
Therefore, their difference is the regret of the agent  when applying strategy $\xi$ in environment $T$. 
Note that, the agent does not know the actual environment $T$ precisely \emph{a priori}. Therefore, the regret of the strategy is considered as the worst-case regret among all possible environments $T\in \T$.

\subsection{Problem Formulation}
After presenting the PK-WTS modeling framework as well as the regret-based performance metric, we are now ready to formulate the problem we solve in this work. 

\begin{problem}[Regret-Based  LTL Planning]\label{pro-1}
Given a possible world represented by PK-WTS $\T$ and an scLTL task $\phi$, 
synthesize a strategy $\xi$ such that i) $ \rho^{T}_\xi\in \L_{\pref}^\phi$ for any $T\in \T$; and ii) $\reg_\T(\xi)$ is minimized. 
\end{problem}

\section{Knowledge-Based Game Arena}\label{sec:KGA}

In this section, we build a knowledge-based game arena to capture all the interaction between the agent and the partially-known environment.

\subsection{Knowledge-Based Game Arena}
Given  PK-WTS  $\T=(X,x_0,\Delta,w,\AP,L)$,  its skeleton system is a WTS 
\[
\mathcal{T}=(X,x_0,\delta_{\mathcal{T}},w,\AP,L),
\]
where  for any $x\in X$, we have 
\[
\delta_{\mathcal{T}}(x)=\bigcup\{ o\in \Delta(x)\}\]
i.e., the successor states of $x$ is defined as the union of all possible successor-patterns. 

To incorporate with the task information,  let $\A_\phi\!=\!(Q,q_0,\Sigma,f,Q_F)$ be the DFA that accepts all good-prefixes of scLTL formula $\phi$. We construct the product system between  $\mathcal{T}$ and $\A_\phi$, denoted by 
\[
\P=\mathcal{T}\otimes \A_\phi=(S,s_0,\delta_\P,w_\P,S_F), 
\]
where the product ``$\otimes$" has been defined in Section \ref{subsec-product} and recall that its state-space is $S=X\times Q$.

However, the state-space of $\P$ is still not sufficient for the purpose of decision-making since the explored knowledges along the trajectory are missing. Therefore, we further incorporate the knowledge-set into the product state-space and explicitly split the movement  choice of the agent and the non-determinism of the environment. This leads to the following \emph{knowledge-based game arena}.

\begin{definition}[Knowledge-Based Game Arena]
Given PK-WTS $\T$, the knowledge-based game arena is a bipartite graph
\[
G=(V=V_a \dot{\cup} V_e,v_0,E),
\]
where
\begin{itemize}
\item 
$V_a\!\subseteq\!X\times Q\times \mathbb{KW}$ is the set of \emph{agent vertices}; 
\item 
$V_e\!\subseteq\! X\times Q\times \mathbb{KW} \times X$ is the set of \emph{environment vertices}; 
\item 
$v_0\!=\! (x_0,q_0,  \K_0 ) \!\in\! V_a$ is the initial (agent) vertex, where  $\K_0$ is the initial knowledge-set defined in \eqref{eq-initialknowledge};
\item 
$E\subseteq V\times V$ is the set of edges defined by:  
for any $v_a=(x_a,q_a,\K_a)\in V_a$ and $v_e=(x_e,q_e,\K_e,\hat{x}_e)\in V_e$, we have
\begin{itemize}
    \item 
    $\< v_a,v_e\>\in E$ whenever
    \begin{enumerate}
        \item[(i)]
        $(x_e,q_e,\K_e)=(x_a,q_a,\K_a)$; and 
        \item[(ii)] 
        $\hat{x}_e\in o_{\K_a}(x_a)$. 
    \end{enumerate} 
    \item 
    $\< v_e,v_a\>\in E$ whenever
    \begin{enumerate}
        \item[(i)]
        $x_a=\hat{x}_e$; and 
        \item[(ii)] 
        $(x_a,q_a)\in \delta_\P(x_e,q_e) $; and 
        \item[(iii)] for every $o\in\Delta(x_a)$, we have
        \begin{equation}{\vspace{3pt}}
            \K_a=\textsf{update}(\K_e,(x_a,o))
        \end{equation}
    \end{enumerate}  
\end{itemize} 
\end{itemize}
\end{definition}

The intuition of the knowledge-based game arena $G$ is explained as follows. The graph is bipartite with two types of vertices: agent vertices from which the agent chooses a feasible successor state to move to and 
environment vertices from which the environment chooses the actual successor-pattern in the possible world. More specifically, for each agent vertex $v_a=(x_a,q_a,\K_a)$, the first component $x_a$ represents its physical state in the system, the second component $q_a$ represents the current DFA state for task $\phi$ and the third component $\K_a$ represents the knowledge-set of the agent obtained along the trajectory.  
At each agent vertex, the agent chooses to move to a successor state. Note that since $x_a$ is the current state, it has been explored and we have $x_a\in X(\K_a)$, i.e., we know that the actual successor states of $x_a$ are $o_{\K_a}(x_a)$. Therefore,  it can move to any environment state $v_e=(x_a,q_a,\K_a, \hat{x})$ by ``remembering" the successor state $\hat{x}\in o_{\K_a}(x_a)$ it chooses. 
Now, at each environment state  $v_e=(x_e,q_e,\K_e,\hat{x}_e)$, the meanings of the first three components are the same as those for agent state. The last component $\hat{x}_e$ denotes the state it is moving to.  
Therefore, $v_e$ can reach agent state $v_a=(x_a,q_a,\K_a)$, where the first two components are just the transition in the product system synchronizing the movements of the WTS and the DFA. Note that we have $x_a=\hat{x}_e$ since the movement has been decided by the agent. However, 
for the last component of knowledge-set $\K_a$, we need to consider the following two cases:
\begin{itemize}
    \item 
    If state $x_a\!=\!\hat{x}_e$ has already been explored, then the agent must observe the same successor-pattern as before. Therefore, the knowledge-set is not updated;  
    \item 
    If state $x_a\!=\!\hat{x}_e$ has not yet been explored, then the new explored knowledge $(x_a, o)\in \kw$ should be added to the knowledge-set $\K_e$. However, since this is the first time the agent visits $x_a$, any possible observations $o\in \Delta(x_a)$ consistent with the prior information are possible. Therefore, the resulting knowledge-set $\K_e$ is non-deterministic. 
\end{itemize}

\begin{remark}
We now discuss the space complexity of the above knowledge-based game arena $G$. 
Let $n\!=\!|X_{un}|$ be the number of the unknown states in $X$. To compute $\mathbb{KW}$, it requires at most $|\mathbb{KW}|\!=\!n!\cdot 2^{n}\cdot |\delta_{\mathcal{T}}|$ space, where we compute $|\delta_{\mathcal{T}}|\!=\!|X|\sum_{x\in X}|\delta_{\mathcal{T}}(x)|$ as the number of all edges in the PK-WTS $\T$.
Thus, to build $G$, it requires $|V|\!=\!n!\cdot 2^n\cdot|X||Q||\delta_{\mathcal{T}}|$ space at most.
\end{remark}

\subsection{Strategies and Plays in the Game Arena}
We call  a finite sequence of vertices 
$\pi=v_0v_1\cdots v_n\in V^*$ a \emph{play} on $G$ if $(v_i,v_{i+1})\!\in\! E$ and we denote by $\textsf{Play}^*(G)$ the set of all finite plays on $G$.
We call $\pi$ a \emph{complete  play} if $\textsf{last}(\pi)\in V_a$, where $\textsf{last}(\pi)$ denotes the last vertex in  $\pi$.  Then for a  complete play 
$\pi=v_0v_1\cdots v_{2n}\in V^*V_a$, 
where $v_{2i}=(x_i, q_i, \K_i ), i=0,\dots, n$, it  induces a path denoted by 
$\pi_{\text{\emph{path}}}= x_0 x_1\cdots x_{n}$ as well as a history 
\begin{equation}
    \pi_{his}= (x_0, o_{\K_0}(x_0))(x_1, o_{\K_1}(x_1)) \cdots 
    (x_{n}, o_{\K_{n}}(x_{n})).   \nonumber
\end{equation}
Note that, in the above, we have $\K_0\subseteq \K_1\subseteq\cdots\subseteq \K_n$ 
and the knowledge-set constructed along history $\pi_{his}$ is exactly $\K_n$. 
On the other hand, for any history 
$\hbar= \kappa_0\kappa_1\cdots\kappa_n\in \kw^*$, there exists a unique complete play in $G$, denoted by $\pi_{\hbar}$,  
such that  its induced history   is $\hbar$.

Since the first two components of $G$ are from the product of $\mathcal{T}$ and $\A_\phi$, 
for any complete play $\pi$, we have $L(\pi_{\text{\emph{path}}})\in \L_{\pref}^\phi$ iff the second component of $\textsf{last}(\pi)$ is an accepting state in the DFA. Therefore, we define 
\begin{equation}
V_F=\{ (x_a,q_a,\K_a)\in V_a \mid q_a\in Q_F \}\nonumber
\end{equation}
the set of accepting vertices representing the satisfaction of the scLTL task. Also, since only  edges from $V_e$ to $V_a$ represent actual movements, we define a weight function for $G$ as 
\begin{equation}\label{eq-originalweight}
  w_G:V\times V\to \RR  
\end{equation}
where for any $v_e\!=\!(x_e,q_e,\K_e, \hat{x}_e)$ and $v_a\!=\!(x_a,q_a,\K_a)$, 
we have $w_G(v_e,v_a)\!=\!w(x_e,x_a)$ and $w_G(v_a,v_e)\!=\!0$. 
The the cost of a play  $\pi\!=\!v_0v_1\cdots v_n\!\in\! V^*$ is defined as 
  $\cost_G(\pi)\!=\!\sum_{i=0}^{n-1}w_G ( v_i,v_{i+1} )$.

A \emph{strategy} for the agent-player is a function 
$\sigma_a\!:\!  V^*V_a \!\to\! V_e \cup\{\stop\}$ 
such that for any $\pi \!\in\! V^*V_a$, either $\< \textsf{last}(\pi),\sigma_a(\pi) \>\!\in\! E$ or $\sigma_a(\pi)\!=\!\stop$.
Analogously, a strategy for the environment-player is a function $\sigma_e\!:\! V^*V_e\!\to\! V_a$ such that for any $\pi\!\in\! V^* V_e$, we have $\< \textsf{last}(\pi),\sigma_e(\pi) \>\!\in\! E$.
We denote by $\Sigma_a(G)$ and $\Sigma_e(G)$ the sets of all strategies for the agent and the environment respectively.
In particular, a strategy $\sigma\in\Sigma_a(G)\cup\Sigma_e(G)$ is said to be \emph{positional} if $\forall \pi,\pi'\!:\!\textsf{last}(\pi)\!=\!\textsf{last}(\pi')\Rightarrow \sigma(\pi)\!=\!\sigma(\pi')$ and we denote by $\Sigma_a^1(G)$ and $\Sigma_e^1(G)$ the corresponding sets of all positional strategies respectively. 
Given strategies $\sigma_a\!\in\!\Sigma_a(G)$ and $\sigma_e\!\in\!\Sigma_e(G)$, the \emph{outcome play} $\pi_{\sigma_a,\sigma_e}$ is the unique sequence 
$v_0v_1\cdots v_n\!\in\! V^* V_a$ s.t.\
\begin{itemize}
    \item 
    $\forall i<n: v_i\in V_a \Rightarrow \sigma_a(v_0v_1\cdots v_i) =v_{i+1}$; and
    \item 
    $\forall i<n: v_i\in V_e\Rightarrow \sigma_e(v_0v_1\cdots v_i) =v_{i+1}$; and
    \item 
    $\sigma_a(v_0v_1\cdots v_n)=\stop$.  
\end{itemize}
By assumption A2, we know that, for two different plays $\pi_1,\pi_2\in V^*V_e$, if $\last(\pi_1)=\last(\pi_2)$, the environment-player should always make the same decision, i.e., $\sigma_e(\pi_1)=\sigma_e(\pi_2)$, since the successor-pattern of the same region is fixed. That is, the environment-player should play a positional strategy $\sigma_e\!\in\!\Sigma_e^1(G)$. 
Furthermore, under assumption A2, for two different environment vertices $v_e\!=\!(x_e,q_e,\K_e,\hat{x}_e),v_e'\!=\!(x_e',q_e',\K_e',\hat{x}_e')\in V_e$, 
if their forth components are the same, i.e., $\hat{x}_e\!=\!\hat{x}_e'$, then the environment-player's decisions, we denote by $\sigma_e(v_e)\!=\!(x,q,\K)$ and $\sigma_e(v_e')\!=\!(x',q',\K')$ with $x\!=\!x'\!=\!\hat{x}_e\!=\!\hat{x}_e'$, should also satisfy $o_{\K}(x)\!=\!o_{\K'}(x)$. To capture the above features of the environment-player's strategy, we define the set of strategies for the environment-player as follows:
\begin{equation}
    \S_e=\left\{\sigma_e\in\Sigma_e^1(G):
    \begin{aligned}
    &\forall v_e,v_e'\!\in\! V_e, \hat{X}(v_e)\!=\!\hat{X}(v_e')\\
    &\Rightarrow o_{\K_1}(\hat{X}(v_e))\!=\!o_{\K_2}(\hat{X}(v_e'))
    \end{aligned}
    \right\}
\end{equation}
where we denote by $\hat{X}(\cdot)$ the forth component of an environment vertex and $\K_1,\K_2$ are the third components of $\sigma_e(\hat{X}(v_e))$ and $\sigma_e(\hat{X}(v_e'))$, respectively.

For the agent-player, we say $\sigma_a$ is winning  if for any $\sigma_e\!\in\!\S_e$, we have $\textsf{last}(\pi_{\sigma_a,\sigma_e})\in V_F$. 
We denote by $\S_a\subseteq \Sigma_a(G)$ the set of all winning strategies. 
Similarly to Definition~\ref{def:regret-in-T}, we can also define the \emph{regret} of an agent-player strategy $\sigma_a\!\in\!\S_a$ in $G$ by 
\begin{equation}\label{eq:reg-G}
    \reg_G(\sigma_a)\!=\!\!\!\max_{\sigma_e\in\S_e}\!\left(\! \cost_G (\pi_{\sigma_a,\sigma_e})-\min_{\sigma_a'\in\S_a}\!\!\!\cost_G(\pi_{\sigma_a',\sigma_e}) \!\!\right)
\end{equation}

Essentially,  an agent-player's strategy $\sigma_a\!\in\! \S_a$ uniquely defines a corresponding strategy  in $\T$, denoted by $\xi_{\sigma_a}\!\in\! \stra_\phi(\T)$ as follows:
for any $\hbar\!\in\! \hist^*(\T)$, we have $ \xi_{\sigma_a}(\hbar)\!=\!\hat{X}(\sigma_a(\pi_\hbar))$. The environment-player's strategy  $\sigma_e\!\in\! \S_e$ essentially corresponds to a possible actual environment $T\!\in\! \T$ since it needs to specify an observation $o\!\in\! \Delta(x)$ for each unexplored $x$, and once $x$ is explored, the observation is fixed based on the construction of $G$.  Since $\cost_G(\cdot)$ is defined only according to its first component, for any play $\pi\!\in\! \play^*(G)$, we have 
$\cost_G(\pi) \!=\!  \cost( \pi_{\text{\emph{path}}} )$.
Therefore, we obtain the following result directly.

\begin{proposition}
Given the PK-WTS $\T$, scLTL task $\phi$, and the knowledge-based game arena $G$, for any strategy $\xi\in\stra_\phi(\T)$, there exists a unique corresponding agent-player strategy $\sigma_a\in\S_a$ such that 
\begin{equation}
    \reg_\T(\xi)=\reg_G(\sigma_a)
\end{equation}
\end{proposition}

With the above result, to solve Problem~\ref{pro-1}, it suffices to find an agent-player strategy in arena $G$ that minimizes the regret defined in \eqref{eq:reg-G}. 
In what follows, we propose an efficient algorithm to synthesize such a strategy.

\section{Game-Based Synthesis Algorithms}\label{sec:algorithm}

In this section, we present the solution of the regret-minimizing game on the knowledge-based game arena.

\subsection{Regret-Minimizing Strategy Synthesis}

To compute the regret for the agent-player, we first define the \emph{best response} for each knowledge-set as follows. Given a knowledge-set $\K\!\in\!\mathbb{KW}$, we denote by $\T_\K$ the refined PK-WTS w.r.t. $\K$ by \eqref{eq:refine}, i.e., 
\begin{equation}
    \T_{\K}=\textsf{refine}(\T,\K)
\end{equation}

\begin{definition}[Best Response]\label{def:br}
Given a PK-WTS $\T$, for each knowledge-set $\K\in\mathbb{KW}$, the agent's best response w.r.t. $\K$ is defined as
\begin{equation}\label{eq:br}
    br(\K)=\min_{T\in\T_\K} \{\cost(\rho):\rho\in\path^*(T),\rho\models\phi\}
\end{equation}
\end{definition}
With a little notation abuse, for each agent vertex $v_a=(x_a,q_a,\K_a)$ in $G$, we define the best response for $v_a$ as the best response w.r.t. $\K_a$, i.e., $br(v_a):=br(\K_a)$.

Intuitively, the best response captures the \emph{optimistic estimation} to the actual environment. 
We denote by $\hat{T}$ the agent's estimate to the actual environment.
That is, for any unknown state $x\!\in\! X_{un}$ in the PK-WTS $\T$, if it has been explored by the knowledge-set $\K$, then the successors of $x$ are updated to $o_\K(x)$, i.e., $\delta_{\hat{T}}(x)\!=\!o_\K(x)$; but if it has not been explored by the knowledge-set $\K$, then the successors of $x$ will be supposed to those that \emph{cooperatively} help the agent finish the task with as less cost as possible, i.e.,
$\delta_{\hat{T}}(x)=\delta_{T'}(x)$, where $T'$ is the compatible environment that minimizes the cost in \eqref{eq:br}.

By definition, we use the \emph{shortest path search} to compute the best response for the knowledge-sets. 
Given a transition system or a graph $T$, we use $\textsf{SP}_T(x,x')$ to denote the shortest path from $x$ to $x'$ in $T$, which can be searched by a standard Djisktra algorithm. Given a set of states $X'\subseteq X$, we use $\textsf{SP}_T(x,X')$ to denote the shortest path from $x$ to $X'$, i.e., $\textsf{SP}_T(x,X')=\min_{x'\in X'}\textsf{SP}_T(x,x')$. Then we can easily compute 
\begin{equation}
    br(\K)=\textsf{SP}_{\mathcal{T}_{\K}\otimes\A_\phi}(q_0,Q_F),
\end{equation}
where $\mathcal{T}_\K$ is the skeleton system of $\T_\K$; we use $\textsf{SP}_G(v,v')$ to denote the shortest path from $v$ to $v'$ in $G$,  and we use 

Apart from the best responses for different knowledge-sets, in theory, we should also compute the actual cost for each complete play $\pi$ in $G$ with $\last(\pi)\!\in\! V_F$. However, since those plays are countless, we need to find some critical plays and record the key costs to compute the regret, instead of listing all the plays and their costs. 
In what follows, we will also use the \emph{shortest path search} to find the critical plays and compute the key costs.

To find the agent's strategy with the minimized regret, by the definition in \eqref{eq:reg-G}, all the strategies in $\S_a$ should be considered. However, due to the same countless issue, it is also difficult to consider all the strategies in $\S_a$. To this end, we will only consider the positional strategies for the agent-player and then prove that is sufficient.

Now, to capture the critical plays and compute the regret of positional agent's strategies, we define a new weight function over the knowledge-based game arena $G$.
First, we define 
\begin{equation}
    E_{\textsf{SP}}\!=\!\left\{(v,v')\!\in\! E\!:\!\exists v_a\in V_F\text{ s.t. }(v,v')\!\in\!\textsf{SP}_G(v_0,v_a)\right\}\notag
\end{equation}
as the set of all edges that are involved in at least one shortest path from the initial vertex $v_0$ to a final vertex $v_a\in V_F$. Then we define the new weight function as
\begin{equation}
    \mu:E\to\RR
\end{equation}
such that 
\begin{itemize}
    \item for any $(v_a,v_e)\in E\cap (V_a\times V_e)$, we have $\mu(v_a,v_e)=0$;
    \item for any $(v_e,v_a)\in E\cap (V_e\times V_a)$, we have
    \begin{itemize}
        \item if $(v_e,v_a)\notin E_{\textsf{SP}}$, then we set $\mu(v_a,v_e)=\infty$;
        \item if $(v_e,v_a)\in E_{\textsf{SP}}$, then we have
        \begin{itemize}
            \item [i)] if $v_a\notin V_F$, we set $\mu(v_a,v_e)=0$;
            \item [ii)] if $v_a\in V_F$, we define
            \begin{equation}
                \mu(v_e,v_a)=\cost_G(\textsf{SP}_G(v_0,v_a))-br(v_a)
            \end{equation}
        \end{itemize}
    \end{itemize}
\end{itemize}

Given the defined new weight function $\mu$, for each play $\pi\!=\!v_0v_1\cdots v_n\!\in\!\play^*(G)$, we define its corresponding cost w.r.t. $\mu$ as
\begin{equation}
    \cost^\mu_G(\pi)=\sum_{i=0}^{n-1}\mu(v_i,v_{i+1})
\end{equation}

The intuition of the above weight function is explained as follows. 
Given an agent-player winning strategy $\sigma_a\!\in\!\S_a$ and an environment-player strategy $\sigma_e\!\in\!\S_e$, we use $\cost^\mu_G(\pi_{\sigma_a,\sigma_e})$ to estimate the regret of strategy $\sigma_a$ after having known that the environment-player plays strategy $\sigma_e$. To be specific, suppose $\pi_{\sigma_a,\sigma_e}\!=\!v_0v_1\cdots v_a$ with $v_a\!=\!(x_a,q_a,\K_a)\in V_F$. 
We use $br(v_a)=br(\K_a)$ to estimate the item $\min_{\sigma_a'\in\S_a}\cost_G(\pi_{\sigma_a',\sigma_e})$ in \eqref{eq:reg-G}, since the knowledge-set $\K_a$ has already recorded the behaviors of the environment's strategy $\sigma_e$.
Moreover, given that the environment-player plays strategy $\sigma_e$ and the agent aims to reach the accepting vertex $v_a$, we use $\cost_G(\textsf{SP}_G(v_0,v_a))$ to record the minimal cost of an agent's strategy that explores the same unknown states with the same order. By the definition of $\mu$, $\cost_G^\mu(\pi_{\sigma_a,\sigma_e})$ essentially captures the minimal regret of the strategy that the agent-player could have played, after having known the environment's strategy $\sigma_e$, to reach the same final vertex $\last(\pi_{\sigma_a,\sigma_e})$ and explore the same unknown states with strategy $\sigma_a$.

Next, with the weight function $\mu$ defined above, we could compute the agent's strategy with minimized regret by solving a \emph{min-max game} over the arena $G$ with weight $\mu$, where the agent-player aims to minimize its cost defined by $\mu$ while the environment-player aims to maximize it.
Now, we summarize the solution to synthesize the strategy that minimizes the regret of the agent when exploring in the partially-known environment as Algorithm~\ref{alg}.

\IncMargin{1em}
\begin{algorithm}[ht]
	\caption{Regret-Minimizing LTL Planning}
	\label{alg}
	\KwIn{PK-WTS $\T$ and scLTL $\phi$}
	\KwOut{Optimal plan $\xi^*$ and minimized $\reg_\T(\xi^*)$}
	
	Construct skeleton-WTS $\mathcal{T}$ and DFA $\A_\phi$ from $\phi$;\\
	Construct knowledge-based game arena $G$;\\
	\ForEach{$(v,v')\!\in\! E$ in $G$} 
    {
    define its weight function $\mu(v,v')$;
    }
    $\sigma_a^*$, $\reg^*\leftarrow \texttt{SolveMinMax}(G,\mu)$;\\
    Obtain strategy $\xi^*$ from $\sigma_a^*$;\\
    \textbf{return} optimal plan $\xi^*$, $\reg^*$\\
	\vspace{8pt}
	\textbf{procedure} $\texttt{SolveMinMax}(G,\mu)$\\
	\tcp{Initialization}
	\ForEach{$v\in V$}{
    \eIf{$v\in V_F$}{
        $W^{(0)}(v)\leftarrow 0$ and $\sigma_a^{(0)}(v)\leftarrow\stop$}
        {$W^{(0)}(v)\leftarrow \infty$}
    }
    \tcp{Value Iteration}
    
    \Repeat{
    $\forall v\in V : W^{(k+1)}(v)=W^{(k)}(v)$
    }{
    \ForEach{$v_e\in V_e$}
    {
    {\small
    $\!\!W^{(k+1)}(v_e)\!\leftarrow\!\!\!\!\!\!\displaystyle\max_{v_a\!\in \succ(v_e)}\!\!\!\left(W^{(k)}(v_a)+\mu(v_e,v_a)\!\right)$
    }
    }
	\ForEach{$v_a\in V_a$}
	{\eIf{$v_a\in V_F$}
	{
	{\small
	$W^{(k+1)}(v_a)\leftarrow 0$ and $\sigma_a^{(k+1)}(v_a)\leftarrow\stop$
	}
	}
	{{\small $\!\!\!\!W^{(k+1)}(v_a)\!\leftarrow\!\!\!\!\!\!\displaystyle\min_{v_e\in\succ(v_a\!)}\!\!\!\left(W^{(k)}(v_e)\!+\!\mu(v_a,v_e)\!\right)$}
	{\footnotesize$\!\!\!\!\!\!\!\!\!\!\!\sigma_a^{(k+1)}(v_a)\!\leftarrow\!\!\!\!\displaystyle\argmin_{\!\!\!v_e\in\succ(v_a\!)}\!\!\!\left(\!W^{(k)}(v_e)\!+\!\mu(v_a,v_e)\!\right)$}
	}
	}
	$k\leftarrow k+1$}
	\textbf{return} $\sigma_a^{(k)},W^{(k)}(v_0)$
\end{algorithm}

\begin{remark}
In the proposed algorithm, we solve the regret-minimizing problem by reducing it to a min-max game with a new weight function, instead of directly adopting a backward value iteration with the original weight function. This is due to the fact that a strategy that minimizes the regret in the whole game does not necessarily minimize the regret in the subgames, which has been shown in \cite{filiot2010iterated} by the corresponding counterexamples.
\end{remark}

\subsection{Properties of the Proposed Algorithm}

In this subsection, we prove the correctness of the proposed algorithm. Before building the soundness and completeness of the our algorithm, we present some important and necessary results showing the properties of the knowledge-based game arena $G$ and the algorithm.

First, since the procedure $\texttt{SolveMinMax}(G,\mu)$ only computes the positional strategies, we present the following result stating that a positional strategy is sufficient to minimize the regret for the agent-player in $G$.

\begin{lemma}\label{lem-positional}
A positional strategy is sufficient to minimize the regret for the agent-player in the game arena $G$, i.e.,
\begin{equation}
    \exists \sigma_a\in\S_a\cap\Sigma_a^1(G),\forall \sigma_a'\in\S_a:\emph{\reg}_G(\sigma_a)\leq\emph{\reg}_G(\sigma_a')\notag
\end{equation}
\end{lemma}

Next, we show that the best response defined in Definition~\ref{def:br} is sufficient to compute the regret for a strategy.

\begin{lemma}\label{lem-br}
Given two strategies $\sigma_a\!\in\!\S_a$ and $\sigma_e\!\in\!\S_e$, 
the regret of strategy $\sigma_a$ against strategy $\sigma_e$ satisfies
\begin{equation}
    \emph{\reg}_G^{\sigma_e}(\sigma_a)\geq\emph{\cost}_G(\pi_{\sigma_a,\sigma_e})-br(\emph{\last}(\pi_{\sigma_a,\sigma_e}))
\end{equation}
In particular, there is an environment strategy $\sigma_e'\in\S_e$ such that
$\emph{\reg}_G^{\sigma_e'}(\sigma_a)\!=\!\emph{\cost}_G(\pi_{\sigma_a,\sigma_e'})\!-\!br(\emph{\last}(\pi_{\sigma_a,\sigma_e'}))$.
\end{lemma}

With the above result, we can use the best response to compute the regret for any agent strategy $\sigma_a\!\in\!\S_a$ by 
$\reg_G(\sigma_a)\!=\!\max_{\sigma_e\in\S_e}\left(\cost(\pi_{\sigma_a,\sigma_e})-br(\last(\pi_{\sigma_a,\sigma_e}))\right)$.

Recall the definition of the weight function $\mu$, which assigns $\infty$ to the edges that are not involved in any a shortest path. Next, we present the result stating that it is sufficient to only consider the strategies that result in shortest outcome plays.

\begin{proposition}\label{prop-positional}
Given a strategy $\sigma_a\!\in\!\S_a$, there is another positional strategy $\sigma_a'\!\in\!\S_a\cap\Sigma_a^1(G)$ such that for any environment strategy $\sigma_e\!\in\!\S_e$, we have
\begin{itemize}
    \item [i)] \emph{$\last(\pi_{\sigma_a',\sigma_e})=\last(\pi_{\sigma_a,\sigma_e})$};
    \item [ii)] \emph{$\cost_G(\pi_{\sigma_a',\sigma_e})=\cost_G\left(\textsf{SP}_G(v_0,\last(\pi_{\sigma_a,\sigma_e})\right))$}.
\end{itemize}
\end{proposition}

Then, we give the following result to show the strategy returned by the procedure $\texttt{SolveMinMax}(G,\mu)$ makes all the outcome plays the shortest paths.

\begin{proposition}\label{prop-return}
Let $(\sigma_a^*,\emph{\reg}^*)$ be the result returned by the procedure \emph{$\texttt{SolveMinMax}(G,\mu)$}. Then, strategy $\sigma_a^*$ is a winning strategy if $\emph{\reg}^*<\infty$. Moreover, for any environment strategy $\sigma_e\!\in\!\S_e$, we have
\begin{equation}
    \emph{\cost}_G(\pi_{\sigma_a^*,\sigma_e})=\emph{\cost}_G\left(\emph{\textsf{SP}}_G(v_0,\emph{\last}(\pi_{\sigma_a^*,\sigma_e}))\right)
\end{equation}
\end{proposition}

Finally, with all the above properties of the knowledge-based game arena $G$ and the iteration of procedure $\texttt{SolveMinMax}(G,\mu)$, we build soundness and completeness of the proposed algorithm as follows.

\begin{theorem}\label{thm}
Let $(\sigma_a^*,\emph{\reg}^*)$ be the result returned by the procedure \emph{$\texttt{SolveMinMax}(G,\mu)$}. The strategy $\sigma_a^*$ minimizes the regret of the agent-player in arena $G$, i.e., 
\begin{equation}
   \emph{\reg}^*\!=\!\emph{\reg}_G(\sigma_a^*)\leq\emph{\reg}_G(\sigma_a), \forall \sigma_a\in\S_a
\end{equation}
Then, Problem \ref{pro-1} is solved by Algorithm \ref{alg}.
\end{theorem}

\begin{remark}\label{remark-complexity}
Now, let us consider the complexity of the proposed algorithm. Since the iteration of the procedure \emph{$\texttt{SolveMinMax}(G,\mu)$} is directly conducted on the original game arena $G$, the space complexity of Algorithm~\ref{alg} is exactly $|V|$. For the time complexity, the iteration of \emph{$\texttt{SolveMinMax}(G,\mu)$} can be finished in at most $|V|$ steps. Therefore, the strategy with minimal regret can be computed in the \emph{polynomial} time in the size of the arena $G$.
Apart from our algorithm, the regret-minimizing strategy synthesis with the same reachability objective is also considered in \cite{filiot2010iterated}, which considers the general two-player game arena rather than the knowledge-based game arena in our work and proposes a graph unfolding approach to unfold the game arena and compute the strategy on the unfolded graph. The algorithm in \cite{filiot2010iterated} is \emph{pseudo-polynomial} which is not only polynomial in the size of the game arena but also in the ratio between the maximal and the minimal weights of the edges in the arena. Due to this issue, if this ratio is very large, it is inevitable that the complexity of the algorithm in \cite{filiot2010iterated} grows rapidly. Compared with \cite{filiot2010iterated}, our algorithm exhibits its merit in overcoming this issue completely.
\end{remark}

\subsection{Other Exploration Strategies}
To compute the exploration strategy with the minimal regret in the partially-known environments $\T$, we reduce it to solving a min-max game with the knowledge-based game arena $G$ and a new weight function $\mu$. As a matter of fact, the other exploration strategies can be also synthesized in the above established framework. Next, we explain the approaches to synthesize two kinds of exploration strategies in the partially-known environments $\T$.

\subsubsection{Worst-Case-based Strategies}

In the Introduction and Section~\ref{subset-regretminmax}, we have introduced the \emph{worst-case-based strategy}, which minimizes the cost function defined in \eqref{eq:worstcase}. Such a strategy can be directly computed by conducting the iteration of min-max game with the knowledge-based game arena $G$ and the original weight function $w$ defined in \eqref{eq-originalweight}.

\subsubsection{Best-Case-based Strategies}

The above worst-case-based strategy essentially characterize the pessimistic estimation to the partially-known environments. Now, we introduce the best-case-based strategy that makes a decision based on the optimistic estimation to the environments. For the agent, it always believes the actual environment as $\mathcal{T}$ and chooses the shortest path in $\mathcal{T}$ to explore the unknown states in the shortest path. After exploring each unknown state, it updates the knowledge-set to $\K$ and still believes the actual environment as $\mathcal{T}_\K$ and repeats the above operation until achieves the LTL task. Such a strategy can also be synthesized by search the shortest path for each vertex in $G$ to the set of final vertices $V_F$.

Intuitively, the regret-based  strategy achieves a reasonable trade-off between the worst-case and best-case strategies by making the trade-off between the actual cost the agent will pay and the potential benefit the agent may obtain for exploring the unknown states. In what follows, we present both simulation and experimental results to show that the regret-based exploration strategy outperforms the other two strategies when implemented in the randomly generated environments.

\section{Simulations and Case Study}\label{sec:case}

In this section, we present both simulation and experimental results to show the effectiveness of the regret-based strategy when exploring the unknown regions in the partially-known environments.

\subsection{Simulations of Randomly Generated Environments}

\begin{figure}
    \centering
    \includegraphics[width=\hsize]{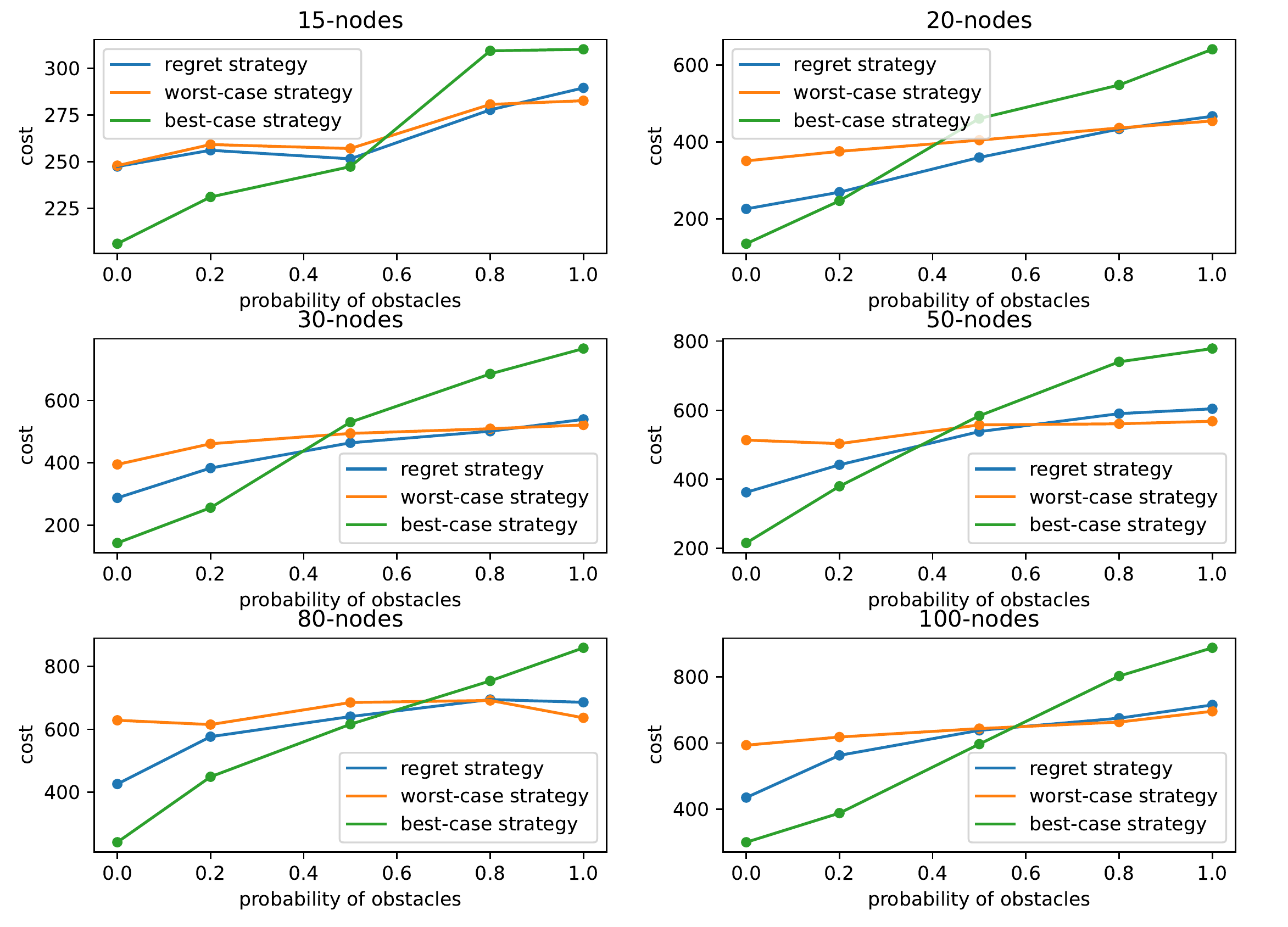}
    \caption{Numerical Simulation Results on Randomly Generated Environments}
    \label{fig-comparison}
\end{figure}

Here we present numerical simulation results to compare the regret-based strategy with the other exploration strategies. We randomly generate the PK-WTS $\T$ with the following parameters: the number of states $|X|$, the number of possible transitions, the minimal and maximal number of successors per state as well as the minimal and maximal transition cost. We set the number of possible transitions to 2, which means that there are two unknown states in the generated PK-WTS.
We set the minimal and maximal number of successors per state to 1 and 2, respectively, and set the minimal and maximal transition cost to 1 and 100, respectively. The agent is assigned with an scLTL task $\lozenge\textsf{target}$ where the atomic proposition $\textsf{target}$ is randomly assigned to several states in the generated PK-WTS. With the above fixed settings, we randomly generated the PK-WTS $\T$ with $|X|\!=\!15,20,30,50,80,100$ repeatedly. Whenever a $\T$ is generated, three different strategies are synthesized based on the algorithms proposed in Section~\ref{sec:algorithm}, including the regret-based strategy, worst-case-based strategy and the best-case-based strategy. Then, for each possible transition in $\T$, we set it with a probability from 0 to 1. Given every PK-WTS $\T$ and a probability of the possible transitions, an actual environment $T\!\in\!\T$ is generated randomly. We apply the three strategies in the same actual environment $T$ and record the corresponding costs the agent pays. For each $|X|$ with each probability, we repeated to randomly generate $\T$ and $T$ for 100 times and compute the average cost for each exploration strategy. Finally, the statistic results are presented in Figure~\ref{fig-comparison}. All codes are available in the project website \href{https://github.com/jnzhaooo/regret}{https://github.com/jnzhaooo/regret}. 

It has been shown in Figure~\ref{fig-comparison} that the average cost for the best-case-based strategy is the most sensitive to the probability of the obstacles. For the regret-based and worst-case-based strategies, the average costs are relatively steady. Furthermore, when the probability of the obstacles are small, the merit of the regret-based strategy is obvious. We can observe that the range of probability that the regret-based strategy outperforms the worst-case-based strategy is obviously larger than the range that the worst-case-based strategy performs better. Therefore, we claim that when the environment is partially-known and the no probabilities are given \emph{a priori}, the regret-based strategy achieves the reasonable trade-off and outperforms other exploration strategies.

Apart from the comparisons between the different exploration strategies, we also compare the space requirement for the different algorithms to solve the regret minimization problem, i.e., our Algorithm~\ref{alg} and the graph-unfolding-based approach in \cite{filiot2010iterated}. 
We set the number of possible transitions to 1 and the minimal and the maximal transition cost to 2 and 5, respectively.
The results are presented in Table~\ref{table}. It is obvious that our algorithm uses less space than that of \cite{filiot2010iterated}, which aligns with the above complexity analysis in Remark~\ref{remark-complexity}.

\begin{table}
  \centering
  \caption{Space requirements for the regret synthesis algorithms}\label{table}
  \setlength{\tabcolsep}{0.8mm}{
  \begin{tabular}{c|cccccc}
     \hline
    $|X|$ & 15 & 20 & 30 & 50 & 80 & 100\\
     \hline
       Ours & 76 & 124 & 133  & 540 & 827 & 2520 \\
       \cite{filiot2010iterated} & {\scriptsize $4.9\!\times\! 10^4$} & {\scriptsize $1.5\!\times\! 10^5$} & {\scriptsize$1.7\!\times\! 10^5$}  & {\scriptsize $2.8\!\times\! 10^6$} & {\scriptsize$6.7\!\times\! 10^6$} & - \\
    \hline
  \end{tabular}}
\end{table}

\subsection{Case Study: A Team of Firefighting Robots}

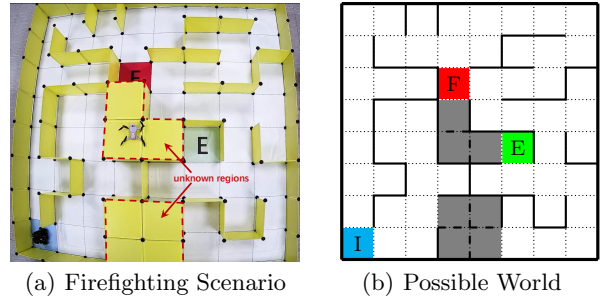
\begin{figure} 
\centering
\subfigure[Firefighting Scenario]{
\includegraphics[scale=0.18]{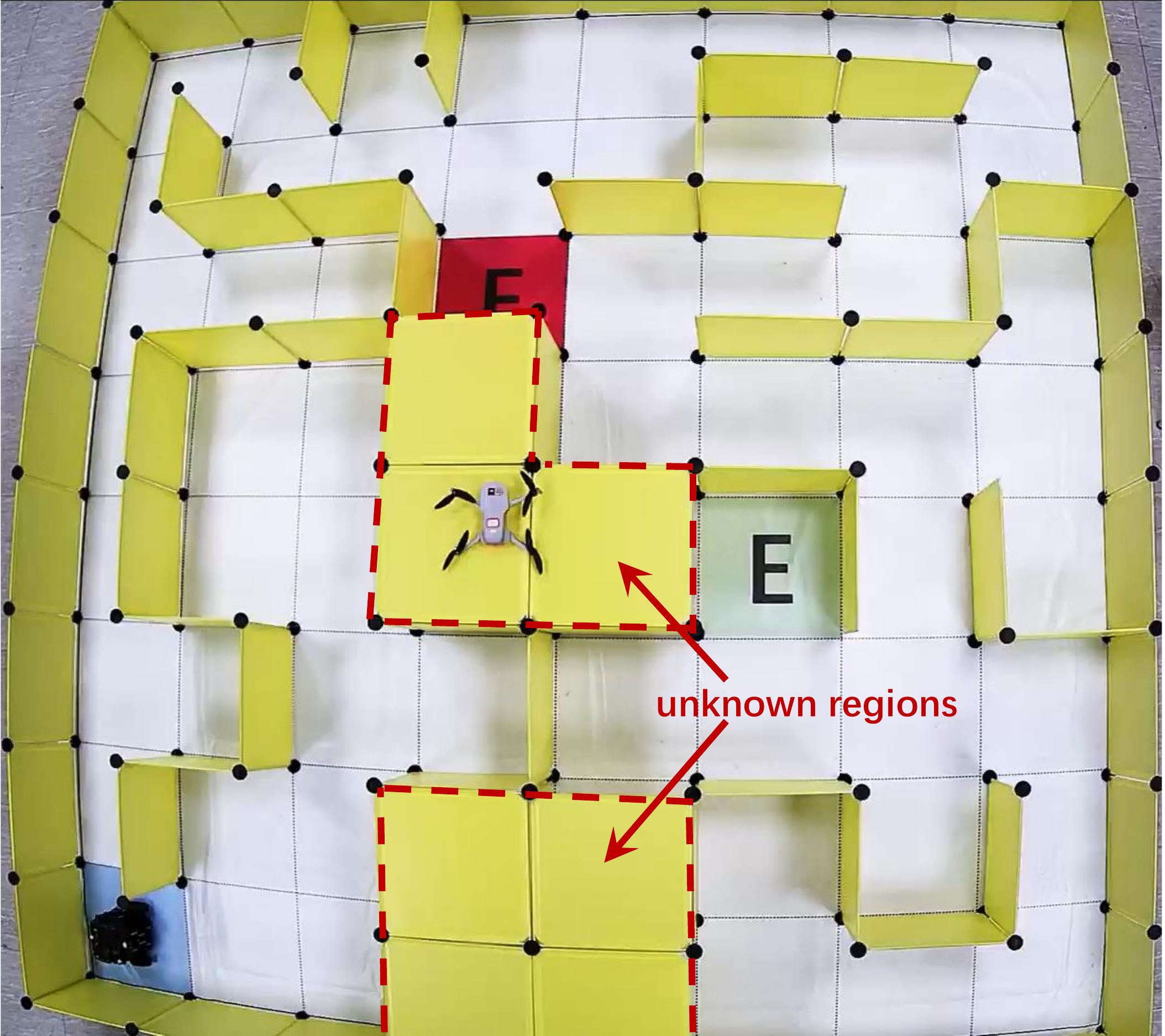}\label{fig:exp}
}
\subfigure[Possible World]{
\newcommand\kkk{2.35}
\definecolor{myyellow}{RGB}{205 192 176}
\begin{tikzpicture}
	\draw[fill=red,draw=white] (3/\kkk,5/\kkk)--(4/\kkk,5/\kkk)--(4/\kkk,6/\kkk)--(3/\kkk,6/\kkk);
	\draw[fill=cyan,draw=white] (0/\kkk,0/\kkk)--(1/\kkk,0/\kkk)--(1/\kkk,1/\kkk)--(0/\kkk,1/\kkk);
	\draw[fill=green,draw=white] (5/\kkk,3/\kkk)--(6/\kkk,3/\kkk)--(6/\kkk,4/\kkk)--(5/\kkk,4/\kkk);
	\draw[fill=gray,draw=white] (3/\kkk,0/\kkk)--(5/\kkk,0/\kkk)--(5/\kkk,2/\kkk)--(3/\kkk,2/\kkk);
	\draw[fill=gray,draw=white] (3/\kkk,3/\kkk)--(5/\kkk,3/\kkk)--(5/\kkk,4/\kkk)--(4/\kkk,4/\kkk)--(4/\kkk,5/\kkk)--(3/\kkk,5/\kkk);

	\foreach \x in {0,2,...,16}
	\draw[densely dotted] (\x/2/\kkk, 0/2/\kkk)--(\x/2/\kkk, 16/2/\kkk);
	\foreach \y in {0,2,...,16}
	\draw[densely dotted] (0/2/\kkk, \y/2/\kkk)--(16/2/\kkk, \y/2/\kkk);
	
	\draw[very thick] (0/\kkk, 0/\kkk)--(0/\kkk, 8/\kkk);
	\draw[very thick] (0/\kkk, 8/\kkk)--(8/\kkk, 8/\kkk);
	\draw[very thick] (8/\kkk, 8/\kkk)--(8/\kkk, 0/\kkk);
	\draw[very thick] (8/\kkk, 0/\kkk)--(0/\kkk, 0/\kkk);
	
	\draw[thick] (1/\kkk, 1/\kkk)--(1/\kkk, 2/\kkk)--(2/\kkk, 2/\kkk)--(2/\kkk, 3/\kkk)--(1/\kkk, 3/\kkk)--(1/\kkk, 5/\kkk)--(3/\kkk, 5/\kkk)--(3/\kkk, 6/\kkk)--(1/\kkk, 6/\kkk)--(1/\kkk, 7/\kkk);
	\draw[thick] (3/\kkk, 5/\kkk)--(3/\kkk, 3/\kkk)--(5/\kkk, 3/\kkk);
	\draw[thick] (2/\kkk, 7/\kkk)--(2/\kkk, 8/\kkk);
	\draw[thick] (3/\kkk, 7/\kkk)--(3/\kkk, 8/\kkk);
	\draw[thick] (4/\kkk, 6/\kkk)--(6/\kkk, 6/\kkk);
	\draw[thick] (5/\kkk, 6/\kkk)--(5/\kkk, 7/\kkk)--(7/\kkk, 7/\kkk);
	\draw[thick] (8/\kkk, 6/\kkk)--(7/\kkk, 6/\kkk)--(7/\kkk, 5/\kkk)--(5/\kkk, 5/\kkk);
	\draw[thick] (4/\kkk, 5/\kkk)--(4/\kkk, 4/\kkk)--(6/\kkk, 4/\kkk)--(6/\kkk, 3/\kkk);
	\draw[thick] (8/\kkk, 3/\kkk)--(7/\kkk, 3/\kkk)--(7/\kkk, 4/\kkk);
	\draw[thick] (7/\kkk, 2/\kkk)--(7/\kkk, 1/\kkk)--(6/\kkk, 1/\kkk)--(6/\kkk, 2/\kkk)--(3/\kkk, 2/\kkk);
	\draw[thick] (3/\kkk, 0/\kkk)--(3/\kkk, 1/\kkk);
	\draw[thick] (5/\kkk, 1/\kkk)--(5/\kkk, 2/\kkk);
	\draw[thick] (4/\kkk, 2/\kkk)--(4/\kkk, 3/\kkk);
	
	\draw[thick, dashed] (4/\kkk, 0/\kkk)--(4/\kkk, 2/\kkk);
	\draw[thick, dashed] (3/\kkk, 1/\kkk)--(5/\kkk, 1/\kkk);
	\draw[thick, dashed] (4/\kkk, 3/\kkk)--(4/\kkk, 4/\kkk)--(3/\kkk, 4/\kkk);
	
	\node[] [xshift=0.5cm/\kkk, yshift=0.5cm/\kkk] (0label) {\scriptsize I};
	\node[] [xshift=5.5cm/\kkk, yshift=3.5cm/\kkk] (0label) {\scriptsize E};
	\node[] [xshift=3.5cm/\kkk, yshift=5.5cm/\kkk] (0label) {\scriptsize F};
	
	\end{tikzpicture}
	
	\label{fig:possibleworld}
}
\caption{Experiment Setting}\label{fig:results}
\end{figure}

In this subsection, we present a case study to illustrate the proposed framework. 
We consider a team of firefighting robots consisting of a ground robot and a UAV. The configuration is shown in Figure~\ref{fig:exp}, where we use ``E" and ``F" to denote \textsf{extinguisher} and \textsf{fire}, respectively.

The firefighting mission   in this district is undertaken by the collaboration of the UAV and the ground robot. Specifically, we assume that the district map is completely unknown to the robotic system initially. When a fire alarm is reported, the UAV takes off first and reconnoiters over the district, which allows the system to obtain some rough information of the distinct and leads to a possible world map, which is shown in Figure~\ref{fig:possibleworld}. More detailed connectivities for some unknown regions in the possible world still remain to be explored by the ground robot. 

In  order to accomplish the firefighting mission, the ground robot needs to first go to the region with \textsf{extinguisher} to get fire-extinguishers and then move to the region with \textsf{fire}. Let $\AP=\{\textsf{fire} , \textsf{extinguisher}\}$. The mission can be described by the following  scLTL formula:
\begin{equation}
 \phi= (\neg \textsf{fire} \ U\  \textsf{extinguisher})\wedge   \lozenge  \textsf{fire} \notag
\end{equation} 

Suppose that, after the reconnaissance, the UAV will get a look down picture of the entire district.
According to the district picture, the system will know the map geometry and the semantics. Specifically, it knows the positions of the \textsf{fire} and \textsf{extinguisher}. However, since some regions are covered by roofs, the connectivities still remain unknown to the system after the reconnaissance. To figure out the (non-)existence of those potential transitions, the ground robot with an onboard camera has to move to the adjacent areas to explore.

Now, the environment is partially-known in the sense that the areas under the roofs are unknown to the robotic system until the ground robot reaches their adjacent regions. Then, based on the possible world model $\T$ and the scLTL $\phi$, we can synthesize the three kinds of exploration strategies by the algorithms in Section~\ref{sec:algorithm} to finish the task $\phi$. Specifically, we consider two compatible environments $T_1,T_2\in\T$ and implement the three strategies in these two environments. 
The results are presented in Figure~\ref{fig:results} and the costs for the three strategies are listed on Table~\ref{table:cost}. We observe that, the regret strategy chooses to explore only one of the unknown regions while the worst-case-based strategy do not explore any regions and the best-case-based strategy explores all the unknown regions. Consequently, for the environment $T_1$, the regret-based strategy saves 14 units cost than the worst-case-based one. Even for the environment $T_2$ where there are obstacles in the both two unknown regions, the regret-based strategy only pays 8 more units than the worst-case-based one. Meanwhile, for both $T_1$ and $T_2$, the regret-based strategy pays less cost than the best-case-based one. Therefore, it has been shown that the regret-based strategy better captures the trade-off between the actual cost and the potential benefit the robot may obtain after exploration. Videos of the above experiments are available at \href{https://youtu.be/lLRT2pLfABA}{https://youtu.be/lLRT2pLfABA}. 

\begin{figure} 
\centering
\subfigure[regret in $T_1$]{
\includegraphics[scale=0.11]{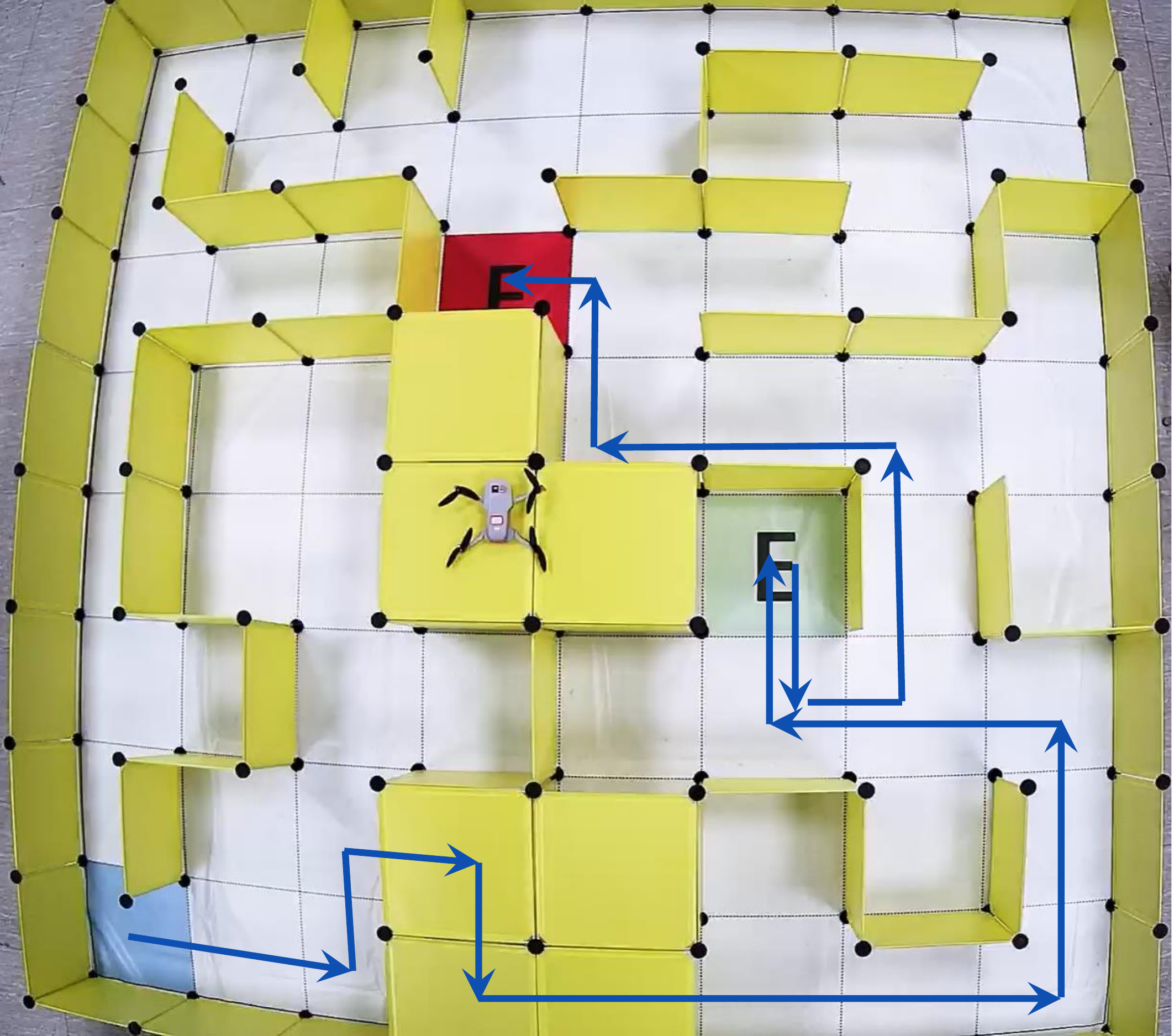}
}
\subfigure[worst in $T_1$]{
\includegraphics[scale=0.11]{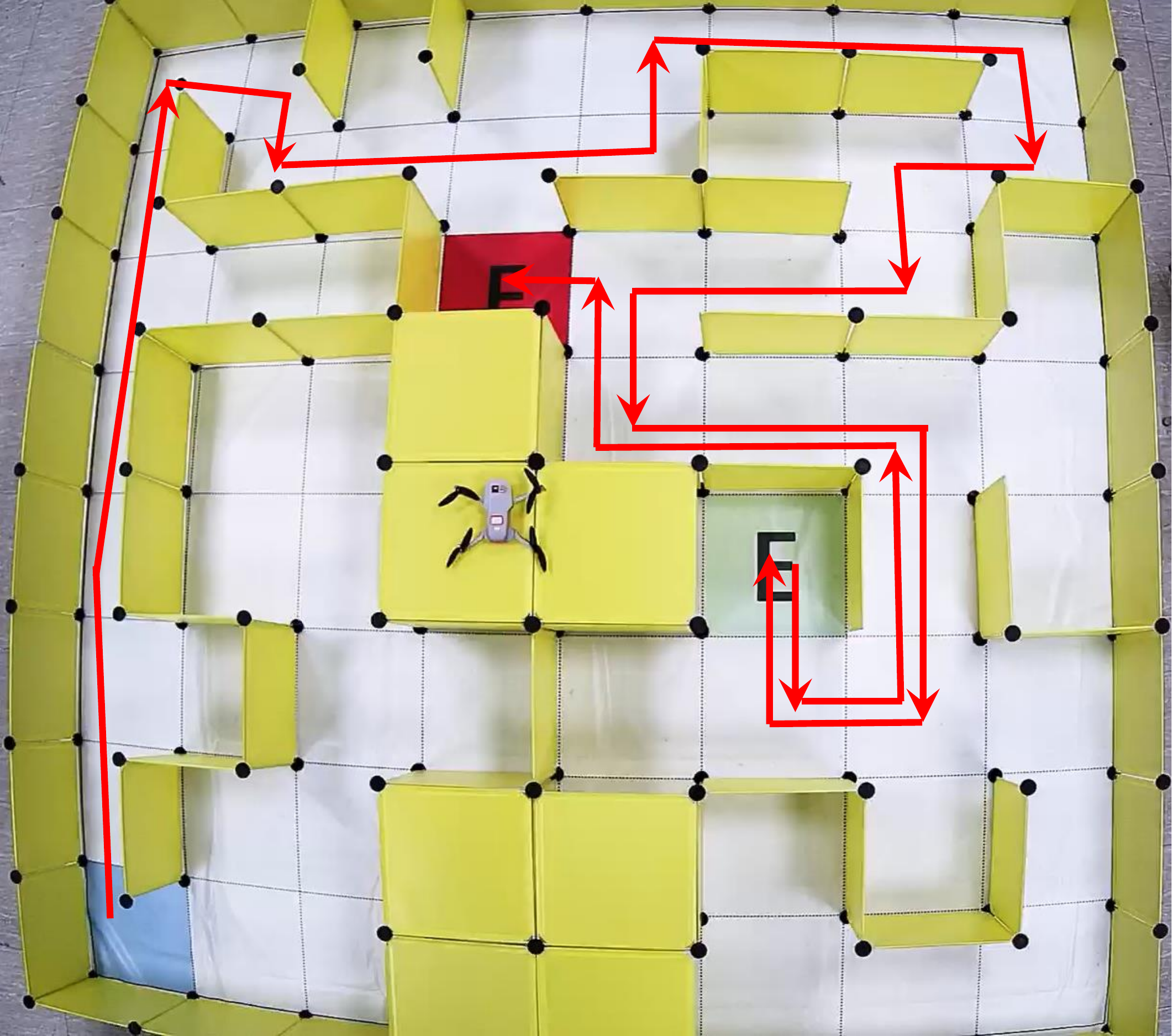}
}
\subfigure[best in $T_1$]{
\includegraphics[scale=0.11]{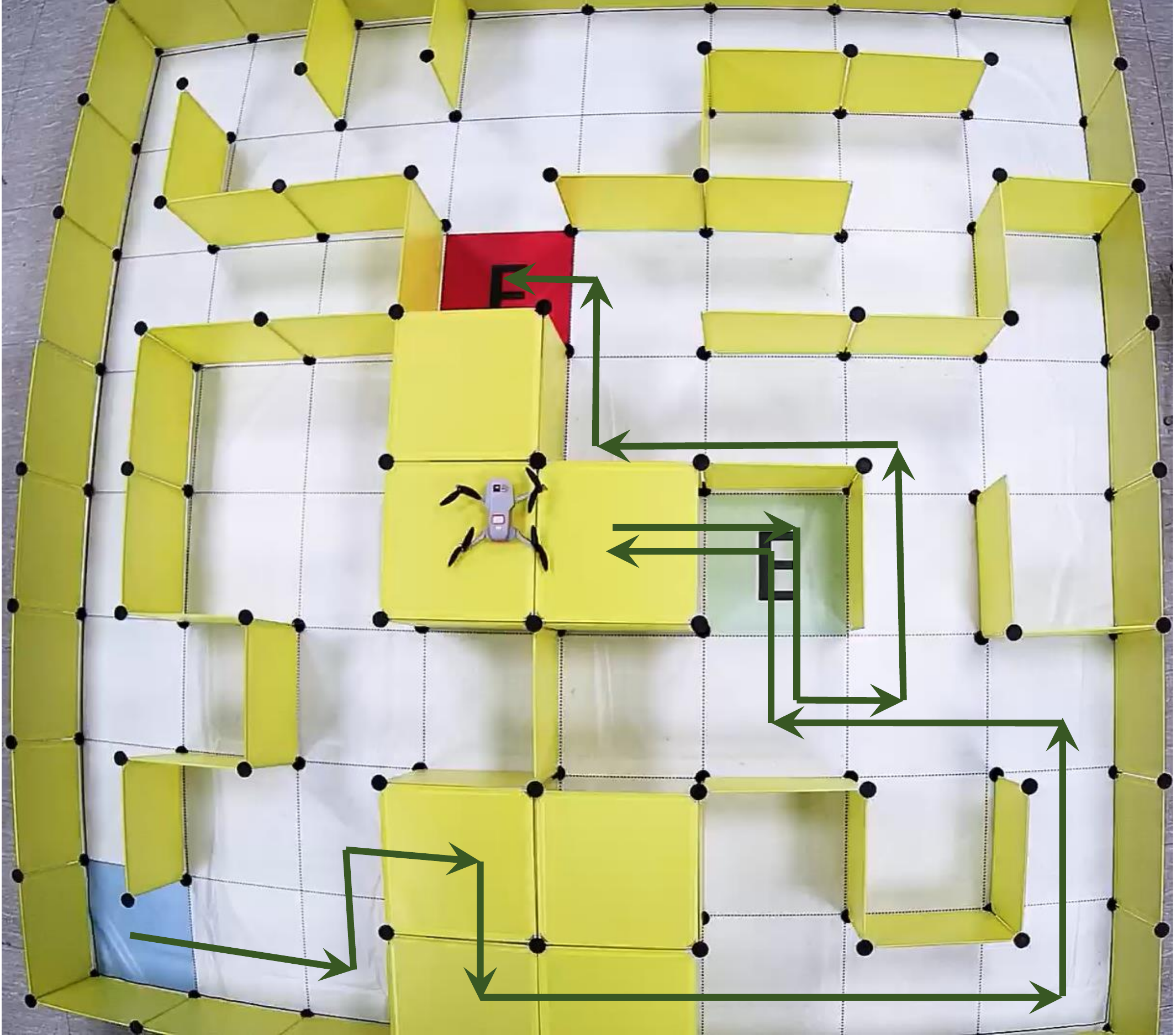}
}
\subfigure[regret in $T_2$]{
\includegraphics[scale=0.11]{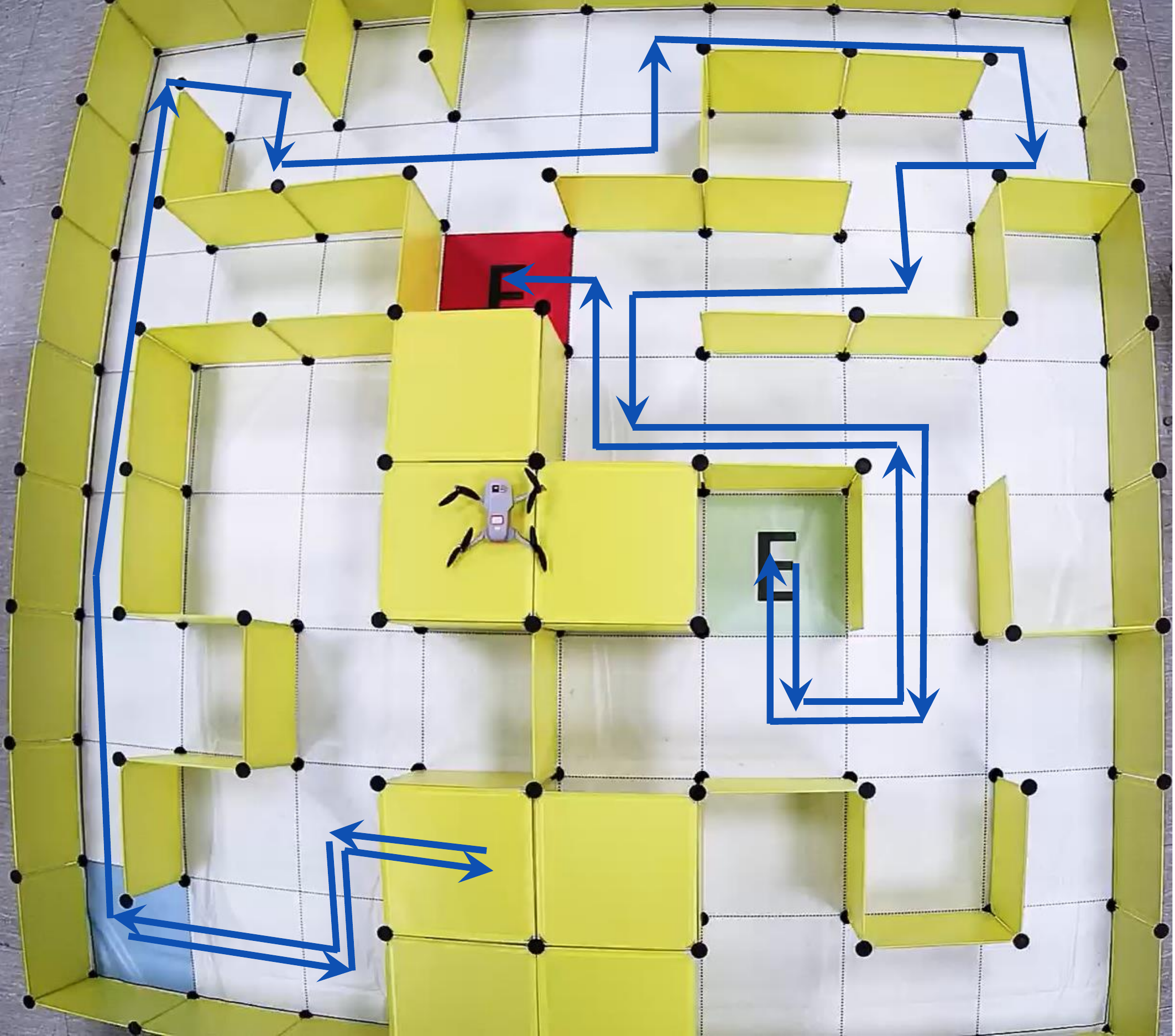}
}
\subfigure[worst in $T_2$]{
\includegraphics[scale=0.11]{fig-worst}
}
\subfigure[best in $T_2$]{
\includegraphics[scale=0.11]{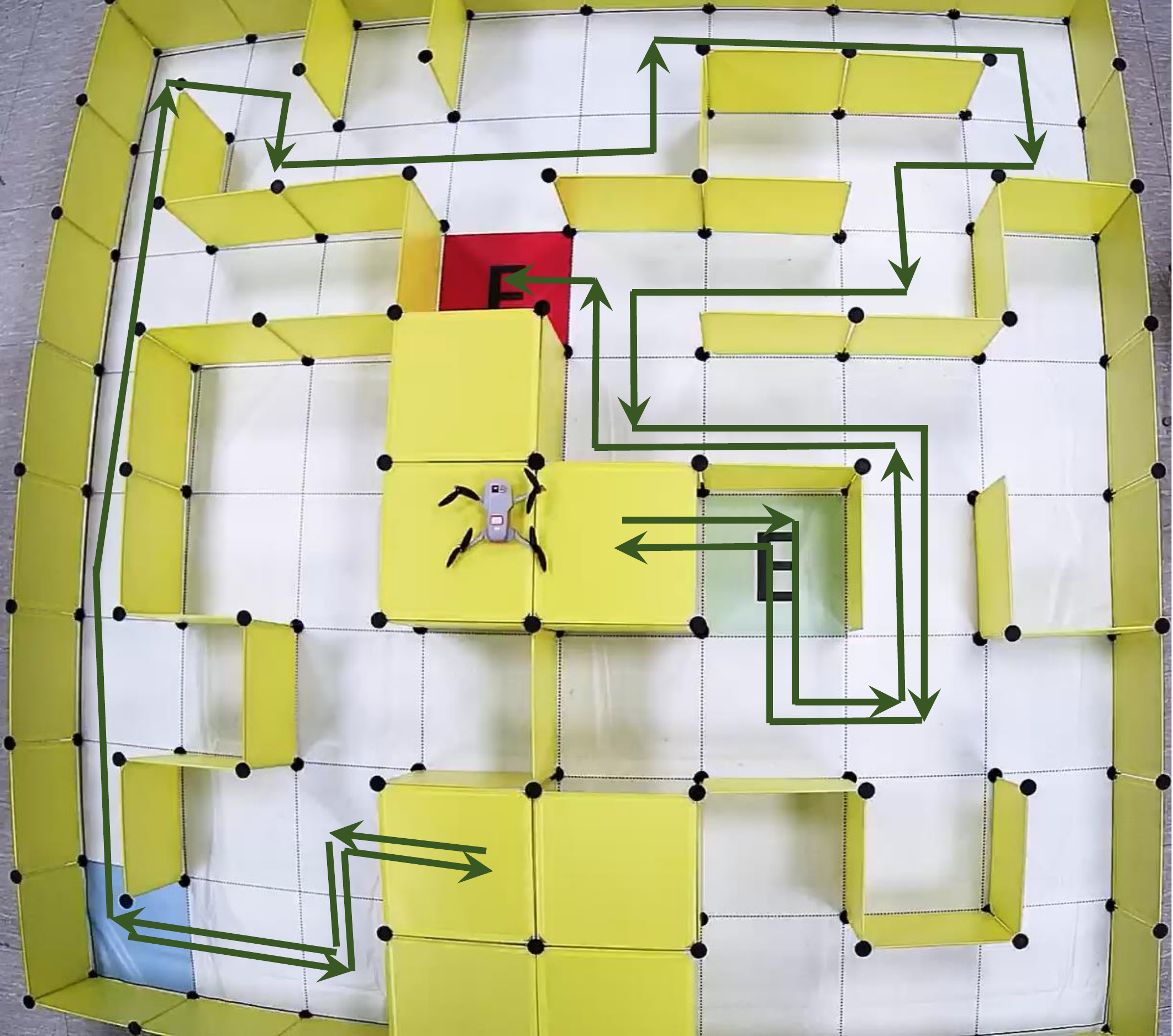}
}
\caption{Experiment Results}
\end{figure} 

\begin{table}
  \centering
  \caption{Costs for the different exploration strategies}\label{table:cost}
  \begin{tabular}{c|cc}
     \hline
    $T$ & $T_1$ & $T_2$ \\
    \hline
    regret-based strategy & 22 & 44\\
    worst-case-based strategy & 36 & 36\\
    best-case-based strategy & 24 & 46\\
    \hline
  \end{tabular}
\end{table}

\section{Conclusions}\label{sec:conclusion}

In this paper, we proposed a new approach for optimal path planning for scLTL specifications under partially-known environments. We adopted the notion of regret to evaluate the trade-off between cost incurred in an actual environment and the potential benefit of exploring unknown regions. A   knowledge-based model was developed to formally describe the partially-known scenario and an effective algorithm was proposed to synthesize an optimal strategy with minimum regret.  In the future, we would like to extend our results to multi-agent systems with general LTL specifications.

\appendix

\section{Proof of Lemma \ref{lem-positional}}
Consider a winning strategy $\sigma_a^F\in\Sigma_a$ for the agent-player in $G$, which means that for any $\sigma_e\!\in\!\S_e$, the outcome play $\pi_{\sigma_a^F,\sigma_e}$ satisfies $\last(\pi_{\sigma_a^F,\sigma_e})\!\in\! V_F$. It is obvious that strategy $\sigma_a^F$ needs at most finite memory since all the outcome plays $\pi_{\sigma_a^F,\sigma_e},\forall \sigma_e\in \S_e$ are finite. Then we show the existence of a memoryless strategy $\sigma_a\in\S_a\cap\Sigma_a^1(G)$ corresponding to $\sigma_a^F$ such that
\begin{equation}\label{eq-A1}
    \forall\sigma_e\in\S_e:\cost_G(\pi_{\sigma_a,\sigma_e})\leq\cost_G(\pi_{\sigma_a^F,\sigma_e}).
\end{equation}
We construct strategy $\sigma_a$ by: for any environment strategy $\sigma_e\!\in\!\S_e$ with $\pi_{\sigma_a^F,\sigma_e}\!=\!v_0v_1\cdots v_n$ being the outcome play and for all $v_k\!\in\! V_a$, we define
\begin{equation}
    \not\exists i\geq k:v_i=v_k\Rightarrow \sigma_a(v_k)=v_{k+1}.
\end{equation}
It is obvious that strategy $\sigma_a$ is well defined. 
By construction, we directly have $\last(\pi_{\sigma_a,\sigma_e})\!\in\! V_F$ for any $\sigma_e\!\in\!\S_e$ and thus $\sigma_a$ is a winning strategy.
Furthermore, for any $\sigma_e\!\in\! \S_e$, the outcome play $\pi_{\sigma_a,\sigma_e}$ only visits the states in $\pi_{\sigma_a^F,\sigma_e}$, since the environment-player plays only positional strategies. Then \eqref{eq-A1} holds since we obtain $\pi_{\sigma_a^F,\sigma_e}$ by removing all cycles in $\pi_{\sigma_a^F,\sigma_e}$ based on the above construction. It directly follows that
\begin{equation}
    \forall\sigma_e\in\S_e:\reg_G^{\sigma_e}(\sigma_a)\leq\reg_G^{\sigma_e}(\sigma_a^F).
\end{equation}
Then we have
\begin{equation}\label{eq-regleq}
    \reg_G(\sigma_a)\leq\reg_G(\sigma_a^F).
\end{equation}
That is, given any winning strategy for the agent-player, we can always find a positional winning strategy making \eqref{eq-regleq} hold. The proof is thus completed.

\section{Proof of Lemma \ref{lem-br}}
Based on the construction of the knowledge-based game arena $G$, we know that, there is a ``one-to-one" correspondence between an environment strategy $\sigma_e\!\in\!\S_e$ and an actual environment $T\!\in\!\T$.
For each actual environment $T$, denote by $\K^T$ the knowledge obtained by the agent after exploring all the unknown states. Formally, we have i) $X(\K^T)\!=\!X$; and ii) for each $(x,o)\!\in\! \K^T$, it holds that $o\!=\!\delta_T(x)$.

Given a play $\pi\!=\!v_0v_1\cdots v_n\in\play^*(G)$, each $v_i\!\in\! V_e$ such that $\succ(v_e)\geq 2$ represents an update of the knowledge, i.e., for such $v_i=(x_i,q_i,\K_i,\hat{x}_i)$ and $v_{i+1}\!=\!(x_{i+1},q_{i+1},\K_{i+1})$, we have $\K_i\subset\K_{i+1}$. Recall that we use $\T_\K\!=\!\textsf{refine}(\T,\K)$ to denote all the possible actual environments that are consistent with knowledge-set $\K$. Then, for each knowledge $\K\in\mathbb{KW}$ and an actual environment $T\!\in\!\T_\K$, there are a sequence of knowledge updates $\K_0,\K_1,\ldots,\K_k$ such that
\begin{equation}
    \K=\K_0\subset\K_1\subset\cdots\subset\K_k=\K^T
\end{equation}
for some $k\!\in\!\NN$.
In particular, if $X(\K)\!=\!X$, we have $\K^T\!=\!\K$ and $k\!=\!0$.

Given the two strategies $\sigma_a$ and $\sigma_e$, denote $\last(\pi_{\sigma_a,\sigma_e})=:(x,q,\K)$. We denote by $T_1,T_2,\ldots,T_m\in\T_\K$ the actual environments that are consistent with knowledge-set $\K$. Then for each environment $T_i$, by the construction of $G$, there is a corresponding environment strategy $\sigma_e^i\!\in\!\S_e$.
Directly, we have $\sigma_e\!\in\!\{\sigma_e^1,\ldots,\sigma_e^m\}$.
By the construction of $G$, we have
\begin{equation}
    \pi_{\sigma_a,\sigma_e^i}=\pi_{\sigma_a,\sigma_e},\forall i=1,\ldots,m.
\end{equation}
Obviously, we have $\cost_G(\pi_{\sigma_a,\sigma_e^i})=\cost_G(\pi_{\sigma_a,\sigma_e})$. By Equation~\eqref{eq:reg-G}, we have
\begin{equation}
    \reg_G^{\sigma_e^i}(\sigma_a)=\cost_G(\pi_{\sigma_a,\sigma_e})-\min_{\sigma_a'\in\S_a}\cost_G(\pi_{\sigma_a',\sigma_e^i})
\end{equation}
For the other hand, based on Definition~\ref{def:br}, we have
\begin{equation}
    br(\K)=\min_{i\in\{1,\ldots,m\}}\min_{\sigma_a'\in\S_a}\cost_G(\pi_{\sigma_a',\sigma_e^i})
\end{equation}
With $br(\K)=br(\last(\pi_{\sigma_a,\sigma_e}))$ and $\sigma_e\in\{\sigma_e^1,\ldots,\sigma_e^m\}$, it directly follows that
\begin{equation}
    \reg_G^{\sigma_e}(\sigma_a)\geq \cost_G(\pi_{\sigma_a,\sigma_e})-br(\last(\pi_{\sigma_a,\sigma_e}))
\end{equation}
The proof is thus completed.

\section{Proof of Proposition \ref{prop-positional}}

Given the strategy $\sigma_a$, consider the outcome play of $\sigma_a$ and an environment strategy $\sigma_e$ which we denote by
\begin{flalign}
&~~~~~~~~~~~~~~~~~~\pi:=\pi_{\sigma_a,\sigma_e}=v_0v_1v_2\cdots v_n\notag\\
&=(x_0,q_0,\K_0)(x_1,q_1,\K_1,\hat{x}_1)(x_2,q_2,\K_2)\cdots(x_n,q_n,\K_n)\notag
\end{flalign}
We consider the environment vertices $v_i\in V_e$ in $\pi$ such that $\succ(v_i)\geq 2$ and denote their index set as
\begin{equation}
    \I_{\pi}=\{i\mid v_i\in (\pi\cap V_e) \wedge \succ(v_i)\geq2\}
\end{equation}
For convenience, we sort set $\I$ as 
\[
\I_{\pi}\!=\!\<\I_{\pi}(1),\ldots,\I_{\pi}(|\I_{\pi}|)\>
\]
Based on the construction of $G$, we know that, along play $\pi$, the agent-player  updates it knowledge-set only \emph{after} these environment  vertices $v_i,i\in\I_{\pi}$. Formally, we have
\begin{itemize}
    \item[i)] $\K_{\I_{\pi}(1)}\subset\K_{\I_{\pi}(1)+1}\subset\K_{\I_{\pi}(2)+1}\subset\cdots\subset\K_{\I_{\pi}(|\I_{\pi}|)+1}$;
    \item[ii)] for any $0\leq j\leq \I(1)$, we have $\K_j=\K_0$; and
    \item[iii)] for any $\I_{\pi}(i)< j \leq \I_{\pi}(i+1)$, we have $\K_j=\K_{\I_{\pi}(i)+1}$.
\end{itemize}

Now, we construct a positional strategy $\sigma_a'\!:\!V_a\!\to\! V_e\cup\{\stop\}$ from the given strategy $\sigma_a$ as follows: for each agent vertex $v_a\!=\!(x_a,q_a,\K_a)\!\in\! V_a$, given any environment strategy $\sigma_e\!\in\!\S_e$, denote $\pi\!:=\!\pi_{\sigma_a,\sigma_e}$ and denote $\K_i$ as the knowledge-set component of $i$-th vertex in $\pi$.
Let $l\!\in\!\I_{\pi}$ be the index such that $\K_l\!=\!\K_a$. Then, we have
\begin{itemize}
    \item if $v_a\neq \last(\pi)$, we define $\sigma_a'$ as follows:
    \begin{itemize}
        \item if $0\!\leq\! l\! \leq \!\I_{\pi}(|\I_{\pi}|)$, then we search the shortest path from $v_a$ to $v_l$ as $\textsf{SP}_G(v_a,v_l)$ and define $\sigma_a'(v_a)\!=\!v_e$ where $v_e$ is the next vertex of $v_a$ in $\textsf{SP}_G(v_a,v_l)$;
    \item if $l\!=\!\I_{\pi}(|\I_{\pi}|)+1$, then we search the shortest path from $v_a$ to $\last(\pi)$ as $\textsf{SP}_G(v_a,\last(\pi))$ and define $\sigma_a'(v_a)\!=\!v_e$ where $v_e$ is the next vertex of $v_a$ in $\textsf{SP}_G(v_a,\last(\pi))$.
    \end{itemize}
    \item if $v_a=\last(\pi)$, we define $\sigma_a'(v_a)=\stop$.
\end{itemize}

Next, we show such function $\sigma_a'\!:\!V_a\!\to\! V_e\cup\{\stop\}$ is a well defined strategy, that is, given any play $\bar{\pi}\!\in\!\play(G)$, only one environment vertex $v_e$ is defined such that $\sigma_a'(\bar{\pi})\!=\!\sigma_a'(\last(\bar{\pi}))\!=\!v_e$. We prove this by contradiction.
Suppose there is a partial play $\bar{\pi}$ with $\last(\bar{\pi})\!=\!\bar{v}_a$ such that both of $\sigma'(\v_a)\!=\!\v_e$ and $\sigma_a'(\v_a)\!=\!\v_e'$ hold, where $\v_e\!\neq\! \v_e'$. Let $\v_i$ be the $i$-th vertex in $\bar{\pi}$ and $\I_{\bar{\pi}}$ be the index set of vertices in $\bar{\pi}$ that have more than two successors. Since the original strategy $\sigma_a\!\in\!\S_a$ is a winning strategy, it is obvious that $\sigma_a$ is well defined on the agent vertex $\v_{\I_{\bar{\pi}}(|\I_{\bar{\pi}}|-1)+1}$. 
On the other hand, based on the definition of $\sigma_a'$, there are two environment strategies $\sigma_e,\sigma_e'\!\in\!\S_e$ such that the outcome plays $\pi_{\sigma_a,\sigma_e}$ and $\pi_{\sigma_a,\sigma_e}$ have the common prefix $\bar{\pi}$ but visit two different environment vertices $v_e$ and $v_e'$, respectively, where $\K(v_e)\!=\!\K(v_e')\!=\!\K(\v_e)\!=\!\K(\v_e')$, $\succ(v_e)\!\geq\! 2$ and $\succ(v_e')\!\geq\! 2$, and we use notation $\K(\cdot)$ to denote the knowledge-set component of a vertex. Based on the construction of the knowledge-based game arena $G$, we know that, in the play $\pi_{\sigma_a,\sigma_e}$, all the vertices between $\v_{\I_{\bar{\pi}}(|\I_{\bar{\pi}}|-1)+1}$ and $v_e$ have the same knowledge-set component and thus all the environment vertices between them have only one successor.
Similarly, we know that, in the play $\pi_{\sigma_a,\sigma_e'}$, all the environment vertices between $\v_{\I_{\bar{\pi}}(|\I_{\bar{\pi}}|-1)+1}$ and $v_e'$ also have only one successor. Since $\sigma_a$ is a well defined strategy, that is, for each agent vertex, there is at most one successor defined by $\sigma_a$, then it should be hold that $v_e\!=\!v_e'$, which contradicts with the fact that $v_e\!\neq\! v_e'$. Therefore,  $\sigma_a'$ is a well defined strategy.

Next, we show the above defined strategy $\sigma_a'$ satisfies that $\last(\pi_{\sigma_a',\sigma_e})\!=\!\last(\pi_{\sigma_a,\sigma_e})$ for any $\sigma_e\!\in\!\S_e$. This is easily guaranteed by the shortest path search from $v_a$ to $\last(\pi_{\sigma_a,\sigma_e})$ and $\sigma_a'(v_a)=\stop$ if $v_a=\last(\pi_{\sigma_a,\sigma_e})$.

Finally, we prove the above defined strategy $\sigma_a'$ makes $\cost_G(\pi_{\sigma_a',\sigma_e})\!=\!\cost_G\left(\textsf{SP}_G(v_0,\last(\pi_{\sigma_a,\sigma_e}))\right)$ for any environment strategy $\sigma_e\!\in\!\S_e$. First, we consider any two partial plays $\pi$ and $\pi'$ such that $\last(\pi)\!=\!\last(\pi')$. We denote by $v_i,v_i'$ the $i$-th vertices in $\pi$ and $\pi'$, respectively, and denote by $\K_i,\K'_i$ the knowledge-set component of $v_i$ and $v_i'$, respectively. 
Let $\I_{\pi}$ and $\I_{\pi'}$ be the index set of vertices in $\pi$ and $\pi'$ that have more than two successors, respectively.
Then, we claim:
\begin{itemize}
    \item[i)] $|\I_{\pi}|=|\I_{\pi'}|$; and
    \item[ii)] for each $0\leq i<|\I_{\pi}|$, we have $v_{\I_{\pi}(i)}\!=\!v'_{\I_{\pi'}(i)}$.
\end{itemize}
To see this, we recall the knowledge-set update rule \eqref{eq-update}. Since the knowledge-set $\K(\last(\pi))\!=\!\K(\last(\pi'))$ is an ordered set, then the knowledge-sets of $\pi$ and $\pi'$ are updated from the initial knowledge-set $\K_0$ to $\K(\last(\pi))$ by adding the same knowledge in sequence. Therefore, $\pi$ and $\pi'$ visited the same environment vertices that have more than two successors before visiting $\last(\pi)$.

Then, consider the two plays $\pi_{\sigma_a',\sigma_e}$ and $\pi_{\sigma_a,\sigma_e}$. Since $\last(\pi_{\sigma_a',\sigma_e})\!=\!\last(\pi_{\sigma_a,\sigma_e})$, we know that, before visiting the final vertex $\last(\pi_{\sigma_a',\sigma_e})$, $\pi_{\sigma_a',\sigma_e}$ and $\pi_{\sigma_a,\sigma_e}$ visited the same environment vertices that have more than two successors in the same sequence. Based on the definition of strategy $\sigma_a'$, the cost of $\pi_{\sigma_a',\sigma_e}$ is minimized since it searched the shortest path between any two environment vertices in $\pi_{\sigma_a,\sigma_e}$ that have more than two successors. The proof is thus completed.

\section{Proof of Proposition~\ref{prop-return}}

By the definition of weight function $\mu$, it directly holds that, for any $\pi\in G$ such that $\last(\pi)\in V_F$, we have
\begin{equation}
 \cost_G^\mu(\pi)\!<\!\infty  ~~\Rightarrow~~ \cost_G(\pi)\!=\!\cost_G(\textsf{SP}_G(v_0,\last(\pi)))\notag
\end{equation}
Then, we first show that the returned $\sigma_a^*$ is a winning strategy if $\reg^*<\infty$. 
It is obvious that $\sigma_a^*$ satisfies
\begin{flalign}
&\reg^* = \max_{\sigma_e\in\S_e}\cost_G^{\mu}(\pi_{\sigma_a^*,\sigma_e})<\infty
\end{flalign}
That is, for any $\sigma_e\in\S_e$, we have $\cost_G^\mu(\pi_{\sigma_a^*,\sigma_e})<\infty$, which means that $\sigma_a^*$ is a winning strategy.
Therefore, it follows that \[
\cost_G(\pi_{\sigma_a^*,\sigma_e})=\cost_G\left(\textsf{SP}_G(v_0,\last(\pi_{\sigma_a^*,\sigma_e}))\right)
\]
The proof is thus completed.

\section{Proof of Theorem~\ref{thm}}

We first characterize the winning strategies in $G$ w.r.t. weight function $\mu$. We define \begin{equation}
    \S_a'=\{\sigma_a'\in\S_a:\cost_G^\mu(\pi_{\sigma_a',\sigma_e})<\infty,\forall \sigma_e\in\S_e\}
\end{equation}
as the set of the winning strategies for the agent-player in $G$ w.r.t. weight function $\mu$. 
Similar to the proof of Proposition~\ref{prop-return}, for any $\sigma_a'\!\in\!\S_a'$ and $\sigma_e\!\in\!\S_e$, we have
\begin{equation}
    \cost_G(\pi_{\sigma_a',\sigma_e})=\cost_G(\textsf{SP}_G(v_0,\last(\pi_{\sigma_a',\sigma_e})))
\end{equation}

By the iteration of the min-max game, it holds that
\begin{equation}\label{eq-regvalue}
    \reg^*=\max_{\sigma_e\in\S_e}\cost_G^{\mu}(\pi_{\sigma_a^*,\sigma_e})
\end{equation}
Moreover, for any $\sigma_a'\in\S_a'$, the returned $\sigma_a^*$ satisfies
\begin{flalign}
    &\max_{\sigma_e\in\S_e}\cost_G^{\mu}(\pi_{\sigma_a^*,\sigma_e})\leq\max_{\sigma_e\in\S_e}\cost_G^{\mu}(\pi_{\sigma_a',\sigma_e})
\end{flalign}
That is, for any strategy $\sigma_a'\in\S_a'$, we have
\begin{flalign}\label{eq-sigma'}
    &\max_{\sigma_e\in\S_e}(\cost_G(\textsf{SP}_G(v_0,\last(\pi_{\sigma_a^*,\sigma_e})))-br(\last(\pi_{\sigma_a^*,\sigma_e})))\notag\\
    &\leq\max_{\sigma_e\in\S_e}(\cost_G(\textsf{SP}_G(v_0,\last(\pi_{\sigma_a',\sigma_e})))\!-\!br(\last(\pi_{\sigma_a',\sigma_e})))
\end{flalign}
By Proposition~\ref{prop-positional}, we know that, for any $\sigma_a\!\in\!\S_a$, there is a strategy $\sigma_a'\!\in\!\S_a'$ such that $\last(\pi_{\sigma_a,\sigma_e})\!=\!\last(\pi_{\sigma_a',\sigma_e})$ for any $\sigma_e\!\in\!\S_e$. Naturally, we have 
\begin{equation}\label{eq-sigma}
    \cost_G(\pi_{\sigma_a,\sigma_e})\geq \cost_G(\pi_{\sigma_a',\sigma_e})
\end{equation}
With \eqref{eq-sigma'} and \eqref{eq-sigma}, it holds that for any $\sigma_a\!\in\!\S_a$, the returned strategy $\sigma_a^*$ satisfies
\begin{flalign}\label{eq-regineq}
    &\max_{\sigma_e\in\S_e}(\cost_G(\textsf{SP}_G(v_0,\last(\pi_{\sigma_a^*,\sigma_e})))-br(\last(\pi_{\sigma_a^*,\sigma_e})))\notag\\
    &\leq\max_{\sigma_e\in\S_e}(\cost_G(\textsf{SP}_G(v_0,\last(\pi_{\sigma_a,\sigma_e})))\!-\!br(\last(\pi_{\sigma_a,\sigma_e})))
\end{flalign}
By Lemma~\ref{lem-br}, for any $\sigma_a\!\in\!\S_a$ and any $\sigma_e\!\in\!\S_e$, we have
\begin{flalign}
&\reg_G^{\sigma_e}(\sigma_a)\geq \cost_G(\pi_{\sigma_a,\sigma_e})-br(\last(\pi_{\sigma_a,\sigma_e}))\notag\\
&~~~~~~~\geq \cost_G\left(\textsf{SP}_G(v_0,\last(\pi_{\sigma_a,\sigma_e}))\right)-br(\last(\pi_{\sigma_a,\sigma_e}))\notag
\end{flalign}
Then it follows that
\begin{flalign}\label{eq-sigma_a}
    \reg_G(\sigma_a)\geq& \max_{\sigma_e\in\S_e}(\cost_G\left(\textsf{SP}_G(v_0,\last(\pi_{\sigma_a,\sigma_e}))\right)\notag\\
    &-br(\last(\pi_{\sigma_a,\sigma_e})))
\end{flalign}
On the other hand, for the returned $\sigma_a^*$, by Lemma~\ref{lem-br} and Proposition~\ref{prop-return}, we have
\begin{flalign}\label{eq-sigma_astar}
    \reg_G(\sigma_a^*)=&\max_{\sigma_e\in\S_e}(\cost_G(\textsf{SP}_G(v_0,\last(\pi_{\sigma_a^*,\sigma_e})))\notag\\
    &-br(\last(\pi_{\sigma_a^*,\sigma_e})))
\end{flalign}
Based on \eqref{eq-regineq}, \eqref{eq-sigma_a}, \eqref{eq-sigma_astar} and \eqref{eq-regvalue}, we have
\begin{equation}
    \reg^*=\reg_G(\sigma_a^*)\leq\reg_G(\sigma_a), \forall \sigma_a\in\S_a
\end{equation}
The proof is thus completed.

\bibliographystyle{plain}
\bibliography{myref}

\end{document}